\pdfoutput=1
\documentclass[5p,number,sort&compress,times,11pt]{elsarticle}
\usepackage{graphicx,pifont,natbib,geometry,fleqn,txfonts}   
\usepackage{amssymb} 
\usepackage{amsmath}
\usepackage{footnote}

\begin{document}
\begin{frontmatter}
\title{Comparison of performance of van der Waals-corrected exchange-correlation functionals for interlayer interaction in graphene and hexagonal boron nitride}

\author[b]{Irina V. Lebedeva\corref{cor}}
\ead{liv\_ira@hotmail.com}
\address[b]{Nano-Bio Spectroscopy Group and ETSF, Universidad del Pa\'is Vasco, CFM CSIC-UPV/EHU, 20018 San Sebastian, Spain
}
\cortext[cor]{Corresponding author}
\author[a]{Alexander V. Lebedev}
\ead{allexandrleb@gmail.com}
\address[a]{Kintech Lab Ltd., 3rd Khoroshevskaya Street 12, Moscow 123298, Russia}
\author[c]{Andrey M. Popov}
\ead{popov-isan@mail.ru}
\address[c]{Institute for Spectroscopy of Russian Academy of Sciences, Troitsk, Moscow 108840, Russia}
\author[a]{Andrey A. Knizhnik}
\ead{kniznik@kintechlab.com}

\begin{abstract}
Exchange-correlation functionals with corrections for van der Waals interactions (PBE-D2, PBE-D3, PBE-D3(BJ), PBE-TS, optPBE-vdW and vdW-DF2) are tested for graphene and hexagonal boron nitride, both in the form of bulk and bilayer. The characteristics of the potential energy surface, such as the barrier to relative sliding of the layers and magnitude of corrugation, and physically measurable properties associated with relative in-plane and out-of-plane motion of the layers including the shear modulus and modulus for axial compression, shear mode frequency and frequency of out-of-plane vibrations are considered. The PBE-D3(BJ) functional gives the best results for the stackings of hexagonal boron nitride and graphite that are known to be ground-state from the experimental studies. However, it fails to describe the order of metastable states of boron nitride in energy. The PBE-D3 and vdW-DF2 functionals, which reproduce this order correctly, are identified as the optimal choice for general studies. The vdW-DF2 functional is preferred for evaluation of the modulus for axial compression and frequency of out-of-plane vibrations, while 
the PBE-D3 functional is somewhat more accurate in calculations of the shear modulus and shear mode frequency. The best description of the latter properties, however, is achieved also using the vdW-DF2 functional combined with consideration of the experimental interlayer distance. In the specific case of graphene, the PBE-D2 functional works very well and can be further improved by adjustment of the parameters.
\end{abstract}
\begin{keyword}
van der Waals interaction \sep density functional theory \sep potential energy surface \sep graphene  \sep hexagonal boron nitride
\end{keyword}
\end{frontmatter}

\section{Introduction}
A number of physical phenomena in two-dimensional bilayers, such as graphene and hexagonal boron nitride, originate from relative displacement of the layers. Relative rotation of the layers gives rise to Moir\'{e} patterns
\cite{Rong1993,Gan2003,Warner2009}, while static translational displacement is manifested through dislocations in stacking of the layers \cite{Alden2013, Brown2012,
Butz2013, Lin2013, Yankowitz2014, Hattendorf2013, San-Jose2014, Lalmi2014, Benameur2015,
Gong2013}. Dynamic phenomena based on relative motion of the layers include atomic-scale slip-stick motion of a flake attached to a STM tip \cite{Dienwiebel2004,Dienwiebel2005,Filippov2008}, rotation-assisted diffusion and drift of a flake \cite{Lebedeva2010,Lebedeva2011a} and self-retracting motion of the layers at their telescopic extension \cite{Zheng2008, Popov2011x}. Based on the link between the relative position of the layers and their electronic properties \cite{Bistritzer2010,Bistritzer2011,Poklonski2013} various types of nanosensors \cite{Poklonski2013,Lam2009,Qian2012,Zheng2012,Paulla2013} can be elaborated. Quantitative description of all of these phenomena and devices depends on the characteristics of the potential surface energy of interlayer interaction, i.e. the dependence of interlayer interaction energy on relative in-plane position of the layers.

Direct investigations of the potential energy surface of layered materials are not accessible to modern experimental methods. Nevetheless, there are a number of physical quantities related to  interlayer interaction that have been measured experimentally and thus provide some insight into characteristics of the potential energy surface. These quantities include among others shear modulus \cite{Bosak2007, Nicklow1972,Grimsditch1983,Seldin1970,Tan2012} and modulus for axial compression \cite{Bosak2007,Nicklow1972,Blakslee1970,Gauster1974,Wada1980,Alzyab1988}, shear mode frequency ($E_{2g}$ mode with adjacent layers sliding rigidly in the opposite
in-plane directions) \cite{Mohr2007, Brillson1971,Tan2012, Boschetto2013,Hanfland1989,Nicklow1972,Boschetto2013}, frequency of relative out-of-plane vibrations ($B_{1g}$ ZO mode with adjacent layers
sliding rigidly towards and away from each other) \cite{Lui2012,Mohr2007, Brillson1971, Alzyab1988,Nicklow1972,Lui2014} and width of dislocations in stacking \cite{Alden2013,Lin2013,Yankowitz2014}. 

Along with these experimental investigations of the properties related to interlayer interaction, significant advances have been achieved in their theoretical description. Semi-empirical potentials for interaction of various two-dimensional layers have been developed, including among others the potentials for graphene \cite{Kolmogorov2005,Lebedeva2011}, boron nitride \cite{Leven2014} and graphene-boron nitride heterostructure \cite{Leven2016}. The registry index surface was introduced to analyze qualitative features of the potential surface of interlayer interaction energy in hexagonal boron nitride bilayer \cite{Marom2010} and graphene-boron nitride heterostructure \cite{Leven2013, Zhao2014, Leven2016}. 
Simple approximations of the potential energy surfaces at a given interlayer distance containing only the first components of Fourier expansions were proposed on the basis of symmetry considerations for bilayer graphene \cite{Ershova2010, Lebedeva2011, Popov2012, Lebedeva2012, Zhou2015}, boron nitride \cite{Lebedev2016,Zhou2015} and graphene-boron nitride heterostructure \cite{ Zhou2015}. These expressions were also extended to take into account the dependence of corrugations of the potential energy surface on the interlayer distance \cite{Reguzzoni2012, Zhou2015}. 

The approximations of the potential energy surface using the first Fourier components include only one energy parameter for graphene \cite{Ershova2010, Lebedeva2011, Popov2012, Lebedeva2012} and boron nitride with the layers aligned in the same direction \cite{Lebedev2016} and two energy parameters for boron nitride with the layers aligned in the opposite directions \cite{Lebedev2016}. This implies that all physical properties characterising the potential energy surface at a given interlayer distance are interrelated. In this way the barrier to relative motion of the layers in graphene bilayer was estimated using the experimental data on the shear mode frequency \cite{Popov2012} and width of dislocations in stacking \cite{Alden2013}. In spite of success of such semi-empirical developments, it should, nevetheless, be kept in mind that they all have been derived or tested on the basis of first-principles calculations and rely on their accuracy. 

The interaction of two-dimensional layers in materials, such as graphene and hexagonal boron nitride, is, however, of long-range van der Waals (vdW) nature and this leads to breakdown of otherwise accurate density functional theory (DFT) methods. Several ways to include description of long-range vdW interactions have been considered. The most straightforward one is just to add the corresponding semi-empirical term \cite{Grimme2006, Grimme2010,Grimme2011, Tkatchenko2009}. Attempts to include the vdW interactions on the fully first-principles footing have been also made through the density-density interaction term describing nonlocal correlations within the long-range correlation function \cite{Dion2004, Lee2010, Klimes2010}. 

In the present paper we address the accuracy of these DFT methods for graphene and hexagonal boron nitride by comparison of the calculation results with the experimentally measurable quantities.
The performance of different DFT methods for description of vdW interlayer interactions has been already addressed in papers \cite{Rego2016,Constantinescu2013,Reguzzoni2012,Zhou2015}. However, paper \cite{Rego2016} was limited to consideration of the properties of graphite related only to changes in the interlayer distance but not to in-plane displacement of the layers. In paper \cite{Zhou2015}, the performance of only two DFT methods, DFT-D2 and vdW-DF2, was compared to the random phase approximation (RPA) data. In papers \cite{Constantinescu2013,Reguzzoni2012}, the authors studied only the characteristics of the potential energy surfaces of hexagonal boron nitride \cite{Constantinescu2013} and graphene \cite{Reguzzoni2012} without relation to any directly measurable quantities. In the present paper, 
we pay attention both to  the properties characterizing out-of-plane motion of the layers, such as the frequency of relative out-of-plane vibrations and modulus for axial compression, which are relevant in systems under external load, as well as the properties associated with in-plane displacement of the layers, such as the shear modulus and shear mode frequency, which are important for lubricity and development of nanoelectromechanical devices based on relative sliding of the layers. Furthermore, the performance of the same exchange-correlation functionals with account of vdW interactions is correlated for two key materials in nanotechnology, graphene and hexagonal boron nitride. The comparison with the \textit{ab initio} methods that adequately describe vdW interactions, such as RPA \cite{Lebegue2010,Zhou2015}, local second-order M{\o}ller-Plesset perturbation theory (LMP2) \cite{Constantinescu2013} and quantum Monte Carlo (QMC) \cite{Mostaani2015,Spanu2009}, is made along with the reference to the experimental data. 

The paper is organized as follows. In section 2 the calculation parameters and considered methods for description of vdW interactions are described. In section 3 the results of calculations for graphene and hexagonal boron nitride are presented and the accuracy of the vdW-corrected exchange-correlation functionals is discussed. Finally the conclusions are summarized.

\section{Methodology}
The DFT calculations have been performed using VASP code \cite{Kresse1996} for the maximum kinetic energy of 600 eV. The projector augmented-wave method \cite{Kresse1999} is used to describe the interaction of valence electrons. The convergence threshold of the self-consistent field is $10^{-8}$ eV.  The rectangular unit cell including 4 atoms in each layer is considered under periodic boundary conditions. The height of the simulation cell is 20~\AA~for bilayers and is equal to the doubled interlayer distance for bulk materials. The Monkhorst-Pack method \cite{Monkhorst1976} is used to perform the integration over the Brillouin zone. The k-point grid is  $24\times 36\times 1$ for bilayers and $24\times 36\times 18$ for bulk (here and below axes $x$ and $y$ are chosen in the armchair and zigzag directions, respectively). Convergence studies carried out previously for graphene bilayer and graphite \cite{Lebedeva2011} showed that these parameters allow to converge the properties of the potential energy surface within the accuracy of 2\%. 

The bond length has been optimized for single-layer graphene and boron nitride using the exchange-correlation functional of Perdew, Burke and Ernzerhof (PBE) \cite{Perdew1996} without inclusion of vdW interactions. The corresponding value for graphene is $l = 1.425$~\AA, close to the experimental data for graphite \cite{Bernal1924, Baskin1955,  Lynch1966, Ludsteck1972, Trucano1975, Zhao1989, Hanfland1989, Bosak2007} and the results of previous calculations \cite{Dion2004,  Kolmogorov2005, Aoki2007, Ershova2010, Lebedeva2011, Reguzzoni2012}. For boron nitride, the optimized bond length is $l = 1.451$~\AA, in agreement with the experimental data for bulk boron nitride \cite{Pease1950, Pease1952, Solozhenko1995,
Solozhenko1997, Solozhenko2001, Paszkowicz2002, Bosak2006, Fuchizaki2008, Yoo1997} and the results of first-principles
calculations \cite{Nag2010, Albe1997, Kern1999, Ohba2001, Janotti2001, Liu2003, Rydberg2003, Tohei2006,
Marini2006, Ooi2006, Serrano2007, Hamdi2010, Constantinescu2013}. The effect of interlayer interaction on
the structure of the layers is neglected \cite{Constantinescu2013}.

The following approaches for description of vdW interactions within DFT are considered: DFT-D2 \cite{Grimme2006}, DFT-D3 \cite{Grimme2010}, DFT-D3(BJ) \cite{Grimme2011}, DFT-TS \cite{Tkatchenko2009}, optPBE-vdW \cite{Klimes2010} and vdW-DF2 \cite{Lee2010}.

In the DFT-D approach \cite{Grimme2006,Grimme2010,Grimme2011}, the total energy $E_{DFT-D}$ is calculated as the sum of the standard Kohn-Sham energy $E_{KS-DFT}$ and the semi-empirical dispersion correction $E_\mathrm{disp}$, so that $E_{DFT-D}=E_{KS-DFT}+E_\mathrm{disp}$. The general form of the dispersion term is 
\begin{equation} \label{eq1}
\begin{split}
E_{\mathrm{disp}} = -\frac{1}{2} \displaystyle \sum_{A \ne B} \displaystyle \sum_{n=6,8,...} 
s_n \frac{C_n^{AB}}{R^n_{AB}} f_\mathrm{dmp,n}(R_{AB}),
\end{split}
\end{equation}
where $C_n^{AB}$ is the $n$-th order dispersion coefficient for the pair of atoms $A$ and $B$, $R_{AB}$ is their internuclear distance,  $s_n$ is the global scaling factor depending on the exchange-correlation functional used and $f_\mathrm{dmp,n}$ is the damping function serving to 
avoid singularities at small $R_{AB}$ and double-counting effects of electron correlation at intermediate distances. 

In the DFT-D2 method \cite{Grimme2006}, only $n=6$ terms are included. The damping function $f_\mathrm{dmp,6}$ has the form of
\begin{equation} \label{eq2}
\begin{split}
f_\mathrm{dmp,6}(R_{AB})=\frac{1}{1+e^{-d\left(R_{AB}/R^{AB}_0-1\right)}},
\end{split}
\end{equation}
where $R^{AB}_0$ is the sum of atomic vdW radii and $d=20$. The $C_6^A$ coefficients for different atoms $A$ are derived on the basis of atomic polarization potentials and static dipole polarizabilities from the PBE0 calculations \cite{Adamo1999}. The geometric mean rule is applied to estimate $C_6^{AB}$ coefficients for pairs of distinct elements $C_6^{AB}=\sqrt{C_6^{A}C_6^{B}}$. The scaling factor $s_6=0.75$ is used for the PBE functional.

In the DFT-D3 approach \cite{Grimme2010}, $n=6$ and $n=8$ terms are considered. The damping function $f_\mathrm{dmp,n}$ has the form of
\begin{equation} \label{eq3}
\begin{split}
f_\mathrm{dmp,n}(R_{AB})=\frac{1}{1+6(R_{AB}/(s_{r,n}R^{AB}_0))^{-\alpha_n}},
\end{split}
\end{equation}
where $s_{r,6}$ is the scaling factor of the cutoff radius $R^{AB}_0$ dependent on the exhange-correlation functional used and parameters $s_{r,8}$, $\alpha_6$ and $\alpha_8$ are fixed at values 1, 14 and 16, respectively. The dispersion coefficients are obtained from the Casimir-Polder formula \cite{Casimir1948} 
\begin{equation} \label{eq4}
\begin{split}
C_6^{AB}=& \frac{3}{\pi}\int_0^\infty d\omega \frac{1}{m}\left[\alpha^{A_mH_p}(i\omega)-\frac{p}{2}\alpha^{H_2}(i\omega)\right]  \\
& \times \frac{1}{k}\left[\alpha^{B_kH_l}(i\omega)-\frac{l}{2}\alpha^{H_2}(i\omega)\right]
\end{split}
\end{equation}
modified to use polarizabilities $\alpha^{M}$ of simple molecules $M=A_mH_p$, $B_kH_l$ and $H_2$ with well-defined electronic structure. In this way, the use of free-atom polarizabilities, which can be strongly influenced by energetically low-lying atomic states, is avoided and the dispersion coefficients dependent on the coordination number of the atoms are introduced. The $C_8^{AB}$  coefficients are computed from $C_6^{AB}$ coefficients on the basis of the power series expansion of dispersion forces (see Ref. \cite{Grimme2010} and references therein). 

In the DFT-D3(BJ) approach \cite{Grimme2011}, the Becke-Jonson (BJ) damping is used 
\begin{equation} \label{eq5}
\begin{split}
f_\mathrm{dmp,n}(R_{AB})=\frac{R^n_{AB}}{R^n_{AB}+(a_1 R^{AB}_0+a_2)^n},
\end{split}
\end{equation}
where $a_1$ and $a_2$ are adjustable parameters. Furthermore, in this method $R^{AB}_0 = \sqrt{C_8^{AB}/C_6^{AB}}$.

The Tkatchenko-Scheffler DFT-TS approach \cite{Tkatchenko2009} is based on the same formal expressions as DFT-D2. However, in this case the dispersion coefficients and damping function depend on the charge density to take into account the effects of local chemical environment of atoms. The  dispersion coefficient $C_6^{A,\mathrm{free}}$ and vdW radius $R^{A,\mathrm{free}}_0$ of free atom $A$ are scaled for atoms in molecules as $C_6^{A}=(V^A_\mathrm{eff}/V^A_\mathrm{free})^2 C_6^{A,\mathrm{free}}$ and $R^{A}_0=(V^A_\mathrm{eff}/V^A_\mathrm{free})^{1/3} R^{A,\mathrm{free}}_0$, where the ratio $V^A_\mathrm{eff}/V^A_\mathrm{free}$ between the effective and free atomic volumes is determined from the Hirshfeld partitioning of the all-electron density. The dispersion coefficients for pairs of atoms are found as
\begin{equation} \label{eq6}
\begin{split}
C_6^{AB}=\frac{2C_6^{A}C_6^{B}}{\frac{\alpha^{B}_0}{\alpha^{A}_0}C_6^{A}+\frac{\alpha^{A}_0}{\alpha^{B}_0}C_6^{B}},
\end{split}
\end{equation}
where $\alpha^{A}_0$ and  $\alpha^{B}_0$ are the static polarizabilities.

Due to the poor performance of the DFT-TS approach for ionic solids in the case of boron nitride we also consider the extension of this method DFT-TS/HI \cite{Bucko2013,Bucko2014}  using iterative Hirshfeld partitioning on the basis of the scheme proposed by Bultinck \cite{Bultinck2007}. In this iterative Hirshfeld algorithm, the neutral reference atoms are replaced by ions with fractional charges determined along with effective atomic volumes at each iteration step. For boron nitride, we also test the DFT-TS+SCS approach \cite{Tkatchenko2012} in which screening of the electrostatic interaction between dipoles by the polarizable surrounding is taken into account. Since we do not observe significant improvement in description of properties of boron nitride when using the DFT-TS/HI and DFT-TS+SCS methods compared to DFT-TS (see section 3), only the DFT-TS approach is considered for graphene bilayer and graphite.

The optPBE-vdW \cite{Klimes2010} and vdW-DF2 \cite{Lee2010} approaches are based on the vdW-DF non-local correlation functional \cite{Dion2004} in which vdW interactions are taken into account through the density-density interaction term.  In the optPBE-vdW functional, the exchange part is optimized for the vdW-DF correlation part on the basis of the S22 benchmark set of weakly interacting dimers and for water clusters \cite{Klimes2010}. In the vdW-DF2 functional \cite{Lee2010}, the use of the more accurate exchange functional is supplemented by application of the large-N asymptote gradient correction in the vdW kernel. An overview of the performance of these approaches for solids in VASP can be found in \cite{Klimes2011}.

\section{Results}

To compare performance of different methods for description of vdW interactions in graphene and hexagonal boron nitride we have carried out calculations for a series of properties related to out-of-plane and in-plane displacements of the layers (Tables~\ref{table:graphene} -- \ref{table:bn_bulk}). The basic properties associated with relative position of the layers are the equilibrium interlayer distance $d_{\mathrm{eq}}$ and binding energy.
The out-of-plane motion of the layers is characterized by the frequency $f_B$  of out-of-plane vibrations of the layers and modulus $C_{33}$ for axial compression. 
As the experimentally measurable properties related to in-plane motion we consider the frequency $f_E$ of in-plane vibrations and shear modulus $C_{44}$. 
We also calculate relative energies of symmetric stackings of the layers at a given interlayer distance. Though such energies cannot be easily measured, there is the experimental evidence for the ground-state and metastable stackings and some indirect estimates are available. 

\begin{table*}
    \caption{Properties of graphene bilayer (equilibrium interlayer distance $d_{\mathrm{eq}}$, binding energy in the AB stacking $E_{\mathrm{AB}}$, magnitude of corrugation of the potential energy surface $\Delta E_{\mathrm{AA}} = E_{\mathrm{AA}} - E_{\mathrm{AB}}$, barrier to relative sliding of the layers $\Delta E_{\mathrm{SP}}= E_{\mathrm{SP}} - E_{\mathrm{AB}}$, shear modulus $C_{44}$, modulus for axial compression $C_{33}$, frequencies of relative in-plane and out-of-plane vibrations $f_E$ and $f_B$, respectively) calculated using different functionals. The data for the experimental interlayer distance $d_{\mathrm{exp}} = 3.34$~\AA~are also given. }
   \renewcommand{\arraystretch}{1.2}
    \resizebox{\textwidth}{!}{
        \begin{tabular}{*{10}{c}}
\hline
Approach & $d_{\mathrm{eq}}$ (\AA) & \begin{tabular}{@{}c@{}} $E_{\mathrm{AB}}$ \\ (meV/atom)  \end{tabular} & \begin{tabular}{@{}c@{}} $\Delta E_{\mathrm{AA}}$ \\ (meV/atom) \end{tabular}  & \begin{tabular}{@{}c@{}} $\Delta E_{\mathrm{SP}}$ \\ (meV/atom) \end{tabular}  & $C_{44}$ (GPa) & $C_{33}$ (GPa) & $f_E$ (cm$^{-1}$) & $f_B$ (cm$^{-1}$) & Ref.  \\\hline
PBE-D2 & 3.256  &  -50.41 &  19.20 & 2.04 & 4.99 & 38.35 & 33.78 & 93.69 & This work  \\\hline
PBE-D3 & 3.530  &  -44.03 &  7.25 & 0.75 & 1.89 & 22.89 & 20.82 & 69.51 & This work  \\\hline
PBE-D3(BJ) & 3.410  &  -48.55 &  11.48 & 1.23 & 3.12 & 30.70 & 26.13 & 81.91 & This work  \\\hline
PBE-TS & 3.360  &  -73.05 & 15.30 & 1.88 & 4.59 & 62.40 & 31.90 & 117.6 & This work  \\\hline
optPBE-vdW & 3.465  &  -59.97 & 9.71 & 1.04 & 2.67 & 31.39 & 23.97 & 82.15 & This work  \\\hline
vdW-DF2 & 3.544  &  -49.11 & 7.33 & 0.77 & 2.07 & 31.73 & 20.84 & 81.68 & This work  \\\hline
PBE-D2 corrected & 3.319  &  -42.674 & 15.62 & 1.67 & 4.09 & 33.35 & 29.79 & 85.09 & This work  \\\hline
PBE ($d=d_{\mathrm{exp}}$) & ~ &  ~ & 14.50  & 1.58 & 3.86 & ~  & 29.35 & ~  & This work  \\\hline
PBE-D2 ($d=d_{\mathrm{exp}}$) & ~ &  ~ & 14.49  & 1.55 & 3.82 & ~  & 29.18 & ~  & This work  \\\hline
PBE-D3 ($d=d_{\mathrm{exp}}$) & ~ &  ~ & 14.38  & 1.60 & 3.91 & ~  & 29.55 & ~  & This work  \\\hline
PBE-D3(BJ) ($d=d_{\mathrm{exp}}$) & ~ &  ~ & 14.57  & 1.58 & 3.91 & ~  & 29.54 & ~  & This work  \\\hline
PBE-TS ($d=d_{\mathrm{exp}}$) & ~ &  ~ & 16.31 & 2.01 & 4.89 & ~  & 33.02 & ~  & This work  \\\hline
optPBE-vdW ($d=d_{\mathrm{exp}}$) & ~ &  ~ & 14.86  & 1.62 & 3.96 & ~  & 29.71 & ~  & This work  \\\hline
vdW-DF2 ($d=d_{\mathrm{exp}}$)  & ~ &  ~ & 14.73  & 1.62 & 3.84 & ~  & 29.29 & ~  & This work  \\\hline
LDA  & 3.33 &  -24.2 & 9.9  & 1.3 & ~ & ~  & ~ & ~  & \cite{Reguzzoni2012} \\\hline
LDA  & 3.33 &  -22.8 & 9.57  & 1.82 & ~ & ~  & ~ & ~  & \cite{Aoki2007} \\\hline
PBE-D2  & 3.31 &  -43.1 & 9.9 & 1.3 & ~ & ~  & ~ & ~  & \cite{Reguzzoni2012} \\\hline
PBE-D2  & 3.25 &  -50.6 & 19.5 & 2.07 & ~ & ~  & ~ & ~  & \cite{Lebedeva2011} \\\hline
PBE-D2  & 3.25 & -50.52 & 12.32 & 1.95 & ~ & 38  & ~ & ~   & \cite{Zhou2015} \\\hline
vdW-DF  & 3.62 &  -49.9 & 5.8 & 0.5 & ~ & ~  & ~ & ~  & \cite{Reguzzoni2012} \\\hline
vdW-DF  & 3.35 &  -29.3 & 18.9 & 1.92 & ~ & ~  & ~ & ~  & \cite{Lebedeva2011} \\\hline
vdW-DF2  & 3.55 & -49.02 & 5.75 & 0.62 & ~ & 30  & ~ & ~   & \cite{Zhou2015} \\\hline
QMC  & $3.43 \pm 0.04$ & $-35.6 \pm 1.6$ & $\sim12.4$ & ~ & ~ & ~  & ~ & $83 \pm 7 $  & \cite{Mostaani2015} \\\hline
RPA  & 3.39 & -91.35 & 8.81 & 1.53 & ~ & 30  & ~ & ~   & \cite{Zhou2015} \\\hline
Exp. 2-3 layers/Cu  & 3.35 & ~ & ~ & ~ & ~ & ~  & ~ & ~  & \cite{Brown2012} \\\hline
Exp. 4 layers/SiC & 3.37 -- $3.46\pm 0.25$ & ~ & ~ & ~ & ~ & ~  & ~ & ~  & \cite{Weng2010} \\\hline
Exp. 9 layers/SiC & $3.370 \pm 0.005$ & ~ & ~ & ~ & ~ & ~  & ~ & ~  & \cite{Varchon2007} \\\hline
Exp. $\le$44 layers/Ni & 3.478 -- 3.490  & ~ & ~ & ~ & ~ & ~  & ~ & ~  & \cite{Zhu2015}\\\hline
Exp. $\le$12 layers/SiO$_2$/Si & $\sim3.7$ & ~ & ~ & ~ & ~ & ~  & ~ & ~  & \cite{Koh2011} \\\hline
Exp.  & ~  & ~ & ~ & ~ & ~ & ~  & $28\pm3$ & ~  & \cite{Boschetto2013} \\\hline
Exp.  & ~  & ~ & ~ & ~ & ~ & ~  & 32 & ~  & \cite{Tan2012} \\\hline
Exp.  & ~  & ~ & ~ & ~ & ~ & ~  & ~ & 80 $\pm$ 2   & \cite{Lui2012} \\\hline
Exp.  & ~  & ~ & ~ & ~ & ~ & ~  & ~ & 81  & \cite{Lui2014} \\\hline
Reference values  & 3.35 & ~ & ~ & ~ & ~ & ~  & 30 & 80.5  & ~\\\hline
\end{tabular}
}
\label{table:graphene}
\end{table*}

\begin{table*}
    \caption{Properties of graphite (equilibrium interlayer distance $d_{\mathrm{eq}}$, binding energy in the AB stacking $E_{\mathrm{AB}}$, magnitude of corrugation of the potential energy surface $\Delta E_{\mathrm{AA}}= E_{\mathrm{AA}} - E_{\mathrm{AB}}$, barrier to relative sliding of the layers $\Delta E_{\mathrm{SP}}= E_{\mathrm{SP}} - E_{\mathrm{AB}}$, shear modulus $C_{44}$, modulus for axial compression $C_{33}$, frequencies of relative in-plane and out-of-plane vibrations $f_E$ and $f_B$, respectively) calculated using different functionals. The data for the experimental interlayer distance $d_{\mathrm{exp}} = 3.34$~\AA~are also given.}
   \renewcommand{\arraystretch}{1.2}
    \resizebox{\textwidth}{!}{
        \begin{tabular}{*{10}{c}}
\hline
Approach & $d_{\mathrm{eq}}$ (\AA) & \begin{tabular}{@{}c@{}} $E_{\mathrm{AB}}$ \\ (meV/atom)  \end{tabular} & \begin{tabular}{@{}c@{}} $\Delta E_{\mathrm{AA}}$ \\ (meV/atom) \end{tabular}  & \begin{tabular}{@{}c@{}} $\Delta E_{\mathrm{SP}}$ \\ (meV/atom) \end{tabular}  & $C_{44}$ (GPa) & $C_{33}$ (GPa) & $f_E$ (cm$^{-1}$) & $f_B$ (cm$^{-1}$) & Ref.  \\\hline
PBE-D2 & 3.223  &  -55.61 &  20.42 & 2.35 & 5.38 & 42.24 & 49.89 & 139.8 & This work  \\\hline
PBE-D3 & 3.485  &  -48.13 &  8.35 & 0.92 & 2.35 & 24.91 & 31.73 & 103.2 & This work  \\\hline
PBE-D3(BJ) & 3.374  &  -53.15 & 12.46 & 1.41 & 3.48 & 32.13 & 39.18 & 119.1 & This work  \\\hline
PBE-TS & 3.332  &  -82.54 &  16.26 & 2.12 & 4.76 & 68.15 & 46.16 & 174.6 & This work  \\\hline
optPBE-vdW & 3.442  &  -64.28 &  10.29 & 1.15 & 2.78 & 32.84 & 35.33 & 119.2 & This work  \\\hline
vdW-DF2 & 3.521  &  -52.37 &  7.89 & 0.87 & 2.22 & 33.30 & 30.69 & 118.7 & This work  \\\hline
PBE-D2 corrected & 3.290 &  -46.91 & 16.50 & 1.87 & 4.29 & 37.13 & 44.10 & 129.69 & This work  \\\hline
PBE ($d=d_{\mathrm{exp}}$) & ~ &  ~ & 13.88  & 1.58 & 3.61 & ~  & 40.16 & ~  & This work  \\\hline
PBE-D2 ($d=d_{\mathrm{exp}}$) & ~ &  ~ & 13.88  & 1.57 & 3.58 & ~  & 39.95 & ~  & This work  \\\hline
PBE-D3 ($d=d_{\mathrm{exp}}$) & ~ &  ~ & 13.74  & 1.61 & 3.66 & ~  & 40.39 & ~  & This work  \\\hline
PBE-D3(BJ) ($d=d_{\mathrm{exp}}$) & ~ &  ~ & 13.96  & 1.59 & 3.63 & ~  & 40.25 & ~  & This work  \\\hline
PBE-TS ($d=d_{\mathrm{exp}}$) & ~ &  ~ & 15.89 & 2.07 & 4.70 & ~  & 45.79 & ~  & This work  \\\hline
optPBE-vdW ($d=d_{\mathrm{exp}}$) & ~ &  ~ & 14.38  & 1.65 & 3.71 & ~  & 40.68 & ~  & This work  \\\hline
vdW-DF2 ($d=d_{\mathrm{exp}}$)  & ~ &  ~ & 14.26  & 1.64 & 3.60 & ~  & 40.10 & ~  & This work  \\\hline
LDA, GGA ($d=d_{\mathrm{exp}}$) & ~ & ~ & 15 & $ \sim$1 & ~ & ~  & ~ & ~ & \cite{Kolmogorov2005} \\\hline
QMC  & $3.426 \pm 0.036$ & $-56 \pm 6$ & ~ & ~ & ~ & ~  & ~ & ~ & \cite{Spanu2009} \\\hline
RPA  & 3.34 & -48 & ~ & ~ & ~ & 36  & ~ & ~ & \cite{Lebegue2010} \\\hline
Exp. 4.2 K &  $3.3360\pm0.0005$  & ~ & ~ & ~ & ~ & ~  & ~ & ~ & \cite{Baskin1955} \\\hline
Exp. 297 K &  $3.3538\pm0.0005$ & ~ & ~ & ~ & ~ & ~  & ~ & ~ & \cite{Baskin1955} \\\hline
Exp. $275\pm2$ K &  3.354 & ~ & ~ & ~ & ~ & ~  & ~ & ~ & \cite{Zhao1989} \\\hline
Exp. $\sim300$ K&  3.356 & ~ & ~ & ~ & ~ & ~  & ~ & ~ & \cite{Trucano1975} \\\hline
Exp. $\sim300$ K &  3.356 & ~ & ~ & ~ & $5.0\pm0.3$ & $38.7\pm0.7$  & ~ & ~ & \cite{Bosak2007} \\\hline
Exp. $\sim300$ K &  $3.353\pm0.002$ & ~ & ~ & ~ & ~ & ~ & $44\pm1$ & ~ & \cite{Hanfland1989} \\\hline
Exp. &  ~  & -52$\pm$5 & ~ & ~ &  ~ &  ~ & ~ & ~ & \cite{Zacharia2004} \\\hline
Exp. &  ~ &  -43$\pm$5 & ~ & ~ &  ~ &  ~ & ~ & ~ & \cite{Girifalco1956} \\\hline
Exp. &  ~   & -35 (+15,-10)  & ~ & ~ &  ~ &  ~ & ~ & ~ & \cite{Benedict1998} \\\hline
Exp. &   ~  & -31$\pm$2 & ~ & ~ &  ~ &  ~ & ~ & ~ & \cite{Liu2012} \\\hline
Exp. &  ~  & ~ & ~ & ~ & $4.6\pm0.2$ & $37.1\pm0.5$  & 45 & $ \sim$130 & \cite{Nicklow1972} \\\hline
Exp. &  ~  & ~ & ~ & ~ & 4.3 &  ~ &  44  & ~ & \cite{Tan2012} \\\hline
Exp. &  ~  & ~ & ~ & ~ &  $5.05\pm0.35$ &  ~ & ~ & ~ & \cite{Grimsditch1983} \\\hline
Exp. &  ~  & ~ & ~ & ~ &  $4.0\pm0.4$ &  ~ & ~ & ~ & \cite{Seldin1970} \\\hline
Exp. &  ~  & ~ & ~ & ~ &  ~ &  $36.5\pm1.0$ & ~ & ~ & \cite{Blakslee1970} \\\hline
Exp. &  ~  & ~ & ~ & ~ &  ~ &  40.7  & ~ & ~ & \cite{Gauster1974} \\\hline
Exp. &  ~  & ~ & ~ & ~ &  ~ &  $36.6\pm0.1$ & ~ & ~ & \cite{Wada1980} \\\hline
Exp. &  ~  & ~ & ~ & ~ &  ~ &  $34.0\pm0.2$ & ~ & 127 & \cite{Alzyab1988} \\\hline
Exp. &  ~  & ~ & ~ & ~ &  ~ &  ~ & 42 & 127 & \cite{Mohr2007, Brillson1971} \\\hline
Exp. &  ~  & ~ & ~ & ~ &  ~ &  ~ & 44 & ~ & \cite{Boschetto2013} \\\hline
Exp. &  ~  & ~ & ~ & ~ &  ~ &  ~ & ~ & 132.3 & \cite{Lui2012} \\\hline
Reference values  & 3.336 & -40.3 & ~ & ~ & 4.59 & 37.3 & 43.5 & 129 & ~\\\hline
\end{tabular}
}
\label{table:graphite}
\end{table*}

\begin{table*}
    \caption{Energies of symmetric stackings of hexagonal boron nitride bilayer (for the layers aligned in the opposite directions: equilibrium interlayer distance $d_{\mathrm{eq}}$, binding energy in the AA' stacking $E_{\mathrm{AA'}}$, relative energy of the AB1' stacking $E_{\mathrm{AB1'}}= E_{\mathrm{AB1'}} - E_{\mathrm{AA'}}$, magnitude of corrugation of the potential energy surfaces $\Delta E_{\mathrm{AB2'}} = E_{\mathrm{AB2'}} - E_{\mathrm{AA'}}$, barriers to relative sliding of the layers $\Delta E_{\mathrm{SP'}}= E_{\mathrm{SP'}} - E_{\mathrm{AA'}}$; for the layers aligned in the same direction: relative energy of the AB stacking $\Delta E_{\mathrm{AB}}=E_{\mathrm{AB}} - E_{\mathrm{AA'}}$, magnitude of corrugation of the potential energy surface $\Delta E_{\mathrm{AA}}= E_{\mathrm{AA}} - E_{\mathrm{AB}}$, barrier to relative sliding of the layers $\Delta E_{\mathrm{SP}}= E_{\mathrm{SP}} - E_{\mathrm{AB}}$) calculated using different functionals. The data for the experimental interlayer distance $d_{\mathrm{exp}} = 3.30$~\AA~are also given.}
   \renewcommand{\arraystretch}{1.2}
    \resizebox{\textwidth}{!}{
        \begin{tabular}{*{10}{c}}
\hline
Approach & $d_{\mathrm{eq}}$ (\AA) & \begin{tabular}{@{}c@{}} $E_{\mathrm{AA'}}$ \\ (meV/atom)  \end{tabular} & \begin{tabular}{@{}c@{}} $E_{\mathrm{AB1'}}$ \\ (meV/atom) \end{tabular}  & \begin{tabular}{@{}c@{}} $\Delta E_{\mathrm{AB2'}}$ \\ (meV/atom) \end{tabular} & \begin{tabular}{@{}c@{}} $\Delta E_{\mathrm{SP'}}$ \\ (meV/atom) \end{tabular}  & \begin{tabular}{@{}c@{}} $\Delta E_{\mathrm{AB}}$ \\ (meV/atom) \end{tabular} & \begin{tabular}{@{}c@{}} $\Delta E_{\mathrm{AA}}$ \\ (meV/atom) \end{tabular}  &\begin{tabular}{@{}c@{}} $\Delta E_{\mathrm{SP}}$ \\ (meV/atom) \end{tabular}  & Ref.  \\\hline
PBE-D2 &  3.120  &  -68.77 &  3.39  & 26.86 & 4.70  & -1.39 & 35.10  & 3.58 & This work  \\\hline
PBE-D3 &  3.438  &  -44.04 &  2.21  & 10.82  & 2.50  & 0.08  & 12.89  & 1.39  & This work  \\\hline
PBE-D3(BJ) &  3.360  &  -46.64 &  1.55  & 13.14 & 2.28  & -0.29  & 15.72  & 1.70  & This work  \\\hline
PBE-TS &  3.368  &  -75.89 &  -0.69  & 10.61  & 0.98  & -1.76 & 15.39 & 1.81  & This work  \\\hline
PBE-TS/HI &  3.474  &  -52.02 &  0.08 & 6.06  & 0.98  & -0.68 & 8.25 & 1.14 & This work  \\\hline
PBE-TS+SCS &  3.411 &  -103.98 &  8.62 & 17.83  & 8.62 & 2.41 & 19.26 & 2.06 & This work  \\\hline
optPBE-vdW &  3.416  &  -59.75 &  1.86  & 11.70  & 2.35 & 0.05  & 13.53  & 1.48  & This work  \\\hline
vdW-DF2 &  3.502  &  -48.42 &  2.36  & 9.553  & 2.54  & 0.58  & 10.26  & 1.13  & This work  \\\hline
PBE ($d=d_{\mathrm{exp}}$)  &  ~  &  ~ &  1.83  & 15.94  & 2.74  & -0.35  & 19.04  & 2.05 & This work  \\\hline
PBE-D2 ($d=d_{\mathrm{exp}}$) & ~  & ~  &  2.15 &  15.91  & 2.90 & -0.32  & 19.24 & 2.03   & This work  \\\hline
PBE-D3 ($d=d_{\mathrm{exp}}$) &  ~  & ~ &  2.92  & 16.47  &  3.46  & -0.05  & 19.77  & 2.10  & This work  \\\hline
PBE-D3(BJ) ($d=d_{\mathrm{exp}}$) &  ~  & ~ & 1.74 &  15.87  & 2.67  & -0.41  & 19.07  & 2.05  & This work  \\\hline
PBE-TS ($d=d_{\mathrm{exp}}$) &  ~  &  ~ &  -1.87  & 13.57  & 0.84  & -2.08  & 18.24  & 2.05  & This work  \\\hline
PBE-TS/HI  ($d=d_{\mathrm{exp}}$)  &  ~  &  ~ &  -0.71 & 13.38  & 1.62  & -1.60 & 17.68 & 2.32 & This work  \\\hline
PBE-TS+SCS ($d=d_{\mathrm{exp}}$)  &  ~  &  ~ &  11.82 & 23.89  & 11.82 & 2.82 & 27.41 & 2.93 & This work  \\\hline
optPBE-vdW ($d=d_{\mathrm{exp}}$) &  ~  &  ~ &  2.29  & 16.64 & 3.11  & -0.15  & 19.59  & 2.13  & This work  \\\hline
vdW-DF2 ($d=d_{\mathrm{exp}}$) &  ~  &  ~ &  3.32  & 17.34  & 3.90  & 0.39  & 19.74  & 2.17  & This work  \\\hline
PBEsol ($d=d_{\mathrm{exp}}$) &  ~  &  ~ &  1.45  & 16.06  & 2.51  & -0.56  & 19.43  & 2.08  & This work  \\\hline
PBEsol-D3 ($d=d_{\mathrm{exp}}$) &  ~  &  ~ &  1.87  & 15.67  & 2.74  & -0.58  & 19.52  & 2.11  & This work  \\\hline
PBEsol-D3(BJ) ($d=d_{\mathrm{exp}}$) &  ~  &  ~ &  1.50  & 16.09  & 2.55  & -0.53  & 19.42 & 2.08  & This work  \\\hline
rPBE ($d=d_{\mathrm{exp}}$) &  ~  &  ~ &  2.01  & 15.78 & 2.84  & -0.23  & 18.70 & 2.02  & This work  \\\hline
rPBE-D3 ($d=d_{\mathrm{exp}}$) &  ~  &  ~ &  2.76  & 15.49 & 3.31  & -0.32  & 19.43 & 2.10 & This work  \\\hline
rPBE-D3(BJ) ($d=d_{\mathrm{exp}}$) &  ~  &  ~ &  2.13  & 16.00 & 2.94 & -0.06  & 18.52 & 1.99 & This work  \\\hline
revPBE ($d=d_{\mathrm{exp}}$) &  ~  &  ~ &  1.97  & 15.85 & 2.82 & -0.25  & 18.79 & 2.02 & This work  \\\hline
revPBE-D3 ($d=d_{\mathrm{exp}}$) &  ~  &  ~ &  2.31  & 15.01 & 2.95 & -0.64  & 19.48 & 2.11 & This work  \\\hline
revPBE-D3(BJ) ($d=d_{\mathrm{exp}}$) &  ~  &  ~ & 1.80  & 15.50 & 2.69 & -0.47  & 18.94 & 2.04 & This work  \\\hline
PBE0 ($d=d_{\mathrm{exp}}$) &  ~  &  ~ & 3.51  & 18.19  & ~ & 0.20  & 21.19 & 2.31 & This work  \\\hline
LMP2 &  3.34 &  ~ &  4.42  & 16.50  & 3.4  & 0.24 & 19.45 & $\sim2.4$  & \cite{Constantinescu2013} \\\hline
PBEsol &  3.33 &  ~ &  3.30  &  14.94 & ~  & 0.30 & 16.92 & $\sim1.8$  & \cite{Constantinescu2013} \\\hline
HF &  3.33 &  ~ &  8.20  &  21.82 & ~  & 1.86 & 23.88 & $\sim2.7$  & \cite{Constantinescu2013} \\\hline
PBE-D & 3.127 &  ~ &  7.90  &  26.16  & ~  & 0.10  & 33.50  & $\sim3.0$ & \cite{Constantinescu2013} \\\hline
PBE0-D & 2.987 &  ~ &  11.38  &  39.76  & ~  & -2.22  & 57.08 & $\sim5.4$ & \cite{Constantinescu2013} \\\hline
PBE ($d=3.34$~\AA) & ~ &  ~ &  3.04  &  14.08  & ~  & 0.26  & 15.96  & $\sim2.0$ & \cite{Constantinescu2013} \\\hline
PBE0 ($d=3.34$~\AA) & ~ &  ~ &  4.52  &  16.16  & ~  & 0.76  & 17.88  & $\sim1.8$ & \cite{Constantinescu2013} \\\hline
PBE-D ($d=3.34$~\AA) & ~ &  ~ & 3.4 &  14.12  & ~  & 0.34  & 16.12  & $\sim1.9$ & \cite{Constantinescu2013} \\\hline
PBE0-D ($d=3.34$~\AA) & ~ &  ~ &  4.88  &  16.20  & ~  & 0.92  & 17.96 & $\sim1.8$ & \cite{Constantinescu2013} \\\hline
PBE-D2 &  3.12 &  -68.53 &  3.24  & 19.11 & $\sim 4.4$ & -1.17 & 22.75 & 3.78 & \cite{Zhou2015} \\\hline
vdW-DF2 &  3.51 &  -50.05 & 2.40   & 7.41  & $> 2.4$  & 0.68 & 7.41 & 1.03 & \cite{Zhou2015} \\\hline
RPA &  3.34 &  -37.62 &  3.57  & 11.04 & $> 3.7$  & 0.76 & 11.40 &   2.06 & \cite{Zhou2015} \\\hline
\end{tabular}
}
\label{table:bn0}
\end{table*}

\begin{table*}
    \caption{Properties of hexagonal boron nitride bilayer (for the layers aligned in the opposite directions: equilibrium interlayer distance $d_{\mathrm{eq}}$, shear modulus $C_{44}$, modulus for axial compression $C_{33}$, frequencies of relative in-plane and out-of-plane vibrations $f_E$ and $f_B$, respectively; for the layers aligned in the same direction: shear modulus $C'_{44}$ and frequency of relative in-plane vibrations $f'_E$) calculated using different functionals. The data for the experimental interlayer distance $d_{\mathrm{exp}} = 3.30$~\AA~are also given.}
   \renewcommand{\arraystretch}{1.2}
    \resizebox{\textwidth}{!}{
        \begin{tabular}{*{9}{c}}
\hline
Approach & $d_{\mathrm{eq}}$ (\AA) & $C_{44}$ (GPa) & $C_{33}$ (GPa) & $f_E$ (cm$^{-1}$) & $f_B$ (cm$^{-1}$) & $C'_{44}$ (GPa) & $f'_E$ (cm$^{-1}$) & Ref.  \\\hline
PBE-D2 &  3.120  &  7.42 &  58.60  & 42.17  & 118.46  & 8.22  & 44.38  & This work  \\\hline
PBE-D3 &  3.438  &  3.58 &  26.18  & 27.90  & 75.42  & 3.47  & 27.44  & This work  \\\hline
PBE-D3(BJ) &  3.360  &  3.97 &  27.16  & 29.73  & 77.71  & 4.14 & 30.33 & This work  \\\hline
PBE-TS &  3.368 &  2.92 &  45.40 & 25.47  & 100.35  & 4.25  & 30.70 & This work  \\\hline
PBE-TS/HI  &  3.474 &  ~ &  26.55  & ~ & 75.56 & ~  & ~  & This work  \\\hline
PBE-TS+SCS  &  3.411 &  7.27 &  42.04  & 39.91  & 95.96  & 5.12  & 33.50 & This work  \\\hline
optPBE-vdW &  3.416  &  3.72 &  29.67  & 28.54 & 80.56 & 3.64  & 28.21  & This work  \\\hline
vdW-DF2 &  3.502  & 3.32 &  32.80  & 26.61  & 83.66  & 2.81  & 24.50 & This work  \\\hline
PBE ($d=d_{\mathrm{exp}}$)  &  ~  &  4.93 &  ~  & 33.40  & ~  & 4.66  & 32.49  & This work  \\\hline
PBE-D2 ($d=d_{\mathrm{exp}}$) &  ~  &  4.74 &  ~  & 32.77  & ~  & 4.90 & 33.30  & This work  \\\hline
PBE-D3 ($d=d_{\mathrm{exp}}$) &  ~  &  5.12 &  ~  & 34.04  & ~  & 5.05  & 33.83 & This work  \\\hline
PBE-D3(BJ) ($d=d_{\mathrm{exp}}$) &  ~  &  4.70 &  ~  & 32.63  & ~  & 4.93  & 33.40  & This work  \\\hline
PBE-TS ($d=d_{\mathrm{exp}}$) &  ~  &  3.18 &  ~  & 26.83 & ~  & 4.74  & 32.78 & This work  \\\hline
PBE-TS+SCS  ($d=d_{\mathrm{exp}}$) &  ~ & 9.72 &   ~  & 46.91  &  ~  & 7.11  & 40.14 & This work  \\\hline
optPBE-vdW ($d=d_{\mathrm{exp}}$) &  ~  &  5.08 &  ~  & 33.93  & ~  & 5.06 & 33.84 & This work  \\\hline
vdW-DF2 ($d=d_{\mathrm{exp}}$) &  ~  &  5.18 & ~   & 34.25  & ~  & 5.56  & 35.50  & This work  \\\hline
PBE-D2  & 3.12 & ~ & 55 & ~ & ~ & ~  & ~    & \cite{Zhou2015} \\\hline
vdW-DF2  & 3.51 & ~ & 24 & ~ & ~ & ~  & ~    & \cite{Zhou2015} \\\hline
RPA  & 3.34 & ~ & 46 & ~ & ~ & ~  & ~   & \cite{Zhou2015} \\\hline
Exp. 10 layers &  $3.25\pm0.10$ & ~ & ~ & ~ & ~ & ~ & ~ & \cite{Warner2010} \\\hline
Exp. $\le20$ layers &  $\sim3.5$ & ~ & ~ & ~ & ~ & ~ & ~ & \cite{Zhi2009} \\\hline
Exp. $\le14$ layers &  $\sim3.5$ & ~ & ~ & ~ & ~ & ~ & ~ & \cite{Nag2010} \\\hline
\end{tabular}
}
\label{table:bn}
\end{table*}

\begin{table*}
    \caption{Energies of symmetric stackings of hexagonal boron nitride bulk (for the layers aligned in the opposite directions: equilibrium interlayer distance $d_{\mathrm{eq}}$, binding energy in the AA' stacking $E_{\mathrm{AA'}}$, relative energy of the AB1' stacking $\Delta E_{\mathrm{AB1'}}= E_{\mathrm{AB1'}} - E_{\mathrm{AA'}}$, magnitude of corrugation of the potential energy surfaces $\Delta E_{\mathrm{AB2'}} = E_{\mathrm{AB2'}} - E_{\mathrm{AA'}}$, barriers to relative sliding of the layers $\Delta E_{\mathrm{SP'}}= E_{\mathrm{SP'}} - E_{\mathrm{AA'}}$; for the layers aligned in the same direction: relative energy of the AB stacking $\Delta E_{\mathrm{AB}}=E_{\mathrm{AB}} - E_{\mathrm{AA'}}$, magnitude of corrugation of the potential energy surface $\Delta E_{\mathrm{AA}}= E_{\mathrm{AA}} - E_{\mathrm{AB}}$, barrier to relative sliding of the layers $\Delta E_{\mathrm{SP}}= E_{\mathrm{SP}} - E_{\mathrm{AB}}$) calculated using different functionals. The data for the experimental interlayer distance $d_{\mathrm{exp}} = 3.30$~\AA~are also given.}
   \renewcommand{\arraystretch}{1.2}
    \resizebox{\textwidth}{!}{
        \begin{tabular}{*{10}{c}}
\hline
Approach & $d_{\mathrm{eq}}$ (\AA) & \begin{tabular}{@{}c@{}} $E_{\mathrm{AA'}}$ \\ (meV/atom)  \end{tabular} & \begin{tabular}{@{}c@{}} $E_{\mathrm{AB1'}}$ \\ (meV/atom) \end{tabular}  & \begin{tabular}{@{}c@{}} $\Delta E_{\mathrm{AB2'}}$ \\ (meV/atom) \end{tabular} & \begin{tabular}{@{}c@{}} $\Delta E_{\mathrm{SP'}}$ \\ (meV/atom) \end{tabular}  & \begin{tabular}{@{}c@{}} $\Delta E_{\mathrm{AB}}$ \\ (meV/atom) \end{tabular} & \begin{tabular}{@{}c@{}} $\Delta E_{\mathrm{AA}}$ \\ (meV/atom) \end{tabular}   & \begin{tabular}{@{}c@{}} $\Delta E_{\mathrm{SP}}$ \\ (meV/atom) \end{tabular}   & Ref.  \\\hline
PBE-D2 &  3.088  &  -76.66 &  4.08  & 30.04  & 5.61 & -1.37 & 39.48 & 4.16 & This work  \\\hline
PBE-D3 &  3.398  &  -48.31 &  2.52  & 12.34  & 2.85  & 0.12  & 14.68 & 1.60 & This work  \\\hline
PBE-D3(BJ) &  3.318  &  -51.35 &  1.78  & 15.12 & 2.67 & -0.28  & 18.05 & 1.99  & This work  \\\hline
PBE-TS &  3.317  & -86.48 &  -2.53  & 13.00  & 0.63  & -1.98  & 16.84  & 1.86 & This work  \\\hline
PBE-TS/HI  &  3.386  & -61.15 &  -0.46  & 10.19  & 1.33  & -1.17  & 13.39  & 1.82 & This work  \\\hline
PBE-TS+SCS  &  3.326  & -72.30 & -2.07  & 13.53  & 0.66  & -1.11  & 15.19  & 1.45 & This work  \\\hline
optPBE-vdW &  3.388  &  -64.61 & 1.89  & 12.82  & 2.49  & -0.02  & 14.96 & 1.66  & This work  \\\hline
vdW-DF2 &  3.478  & -52.09 &  2.36  & 10.30  & 2.61 & 0.52 & 11.25 & 1.25 & This work  \\\hline
PBE ($d=d_{\mathrm{exp}}$)  &  ~  &  ~ &  1.95  & 16.10  & 2.87 & -0.24 & 19.12  & 2.10  & This work  \\\hline
PBE-D2 ($d=d_{\mathrm{exp}}$) &  ~  & ~ &  2.28  & 16.07  & 3.03  & -0.22  & 19.33 & 2.09 & This work  \\\hline
PBE-D3 ($d=d_{\mathrm{exp}}$) &  ~  & ~ &  3.04  & 16.63  & 3.60 & 0.05 & 19.86 & 2.16 & This work  \\\hline
PBE-D3(BJ) ($d=d_{\mathrm{exp}}$) &  ~  & ~ &  1.86  & 16.02  & 2.81  & -0.31  & 19.15 & 2.11  & This work  \\\hline
PBE-TS ($d=d_{\mathrm{exp}}$) &  ~  & ~ &  -2.82  & 13.91  & 0.63  & -2.02  & 17.60 & 1.907  & This work  \\\hline
PBE-TS/HI  ($d=d_{\mathrm{exp}}$) &  ~  & ~ &  -0.66  & 14.82  & 1.71  & -1.55  & 19.08  & 2.36 & This work  \\\hline
PBE-TS+SCS  ($d=d_{\mathrm{exp}}$) &  ~  & ~ &  -2.61  & 14.58  & 0.63 & -1.26  & 16.19  & 1.52 & This work  \\\hline
optPBE-vdW ($d=d_{\mathrm{exp}}$) &  ~  & ~ &  2.26  & 16.79  & 3.15  & -0.15  & 19.83 & 2.20  & This work  \\\hline
vdW-DF2 ($d=d_{\mathrm{exp}}$) &  ~  &  ~ &  3.24  & 17.51  & 3.90  & 0.37  & 20.02 & 2.25  & This work  \\\hline
PBEsol ($d=d_{\mathrm{exp}}$) &  ~  &  ~ & 1.60  & 16.20  & 2.67  & -0.44  & 19.48 & 2.14 & This work  \\\hline
PBEsol-D3 ($d=d_{\mathrm{exp}}$) &  ~  &  ~ & 2.02  & 15.80  & 2.89  & -0.46  & 19.57 & 2.17 & This work  \\\hline
PBEsol-D3(BJ) ($d=d_{\mathrm{exp}}$) &  ~  &  ~ & 1.65  & 16.23  & 2.70  & -0.41  & 19.47 & 2.14 & This work  \\\hline
rPBE ($d=d_{\mathrm{exp}}$) &  ~  &  ~ & 2.11  & 15.96  & 2.96  & -0.14  & 18.82 & 2.07 & This work  \\\hline
rPBE-D3 ($d=d_{\mathrm{exp}}$) &  ~  &  ~ & 2.86  & 15.67  & 3.42 & -0.23  & 19.55 & 2.15 & This work  \\\hline
rPBE-D3(BJ) ($d=d_{\mathrm{exp}}$) &  ~  &  ~ & 2.23  & 16.18  & 3.06 & 0.03  & 18.64 & 2.05 & This work  \\\hline
revPBE ($d=d_{\mathrm{exp}}$) &  ~  &  ~ & 2.07  & 16.02  & 2.95 & -0.16  & 18.90 & 2.08 & This work  \\\hline
revPBE-D3 ($d=d_{\mathrm{exp}}$) &  ~  &  ~ & 2.42 & 15.18  & 3.07 & -0.55  & 19.59 & 2.16 & This work  \\\hline
revPBE-D3(BJ) ($d=d_{\mathrm{exp}}$) &  ~  &  ~ & 1.91 & 15.67  & 2.81 & -0.38  & 19.05 & 2.09 & This work  \\\hline
LMP2 &  3.34 &  ~ &  3.76  & 16.14  & ~  & 0.44 & 19.45 &  $\sim2.3$ & \cite{Constantinescu2013} \\\hline
PBEsol &  3.33 &  ~ &  3.07  &  14.57 & ~  & 0.34 & 16.42 &  $\sim1.8$ & \cite{Constantinescu2013} \\\hline
HF &  3.33 &  ~ &  7.71  &  21.64 & ~  & 1.87 & 23.40 &  $\sim2.7$ & \cite{Constantinescu2013} \\\hline
PBE-D & 2.972 &  ~ &  7.55  &  38.99  & ~  & -3.20  & 57.50  & $\sim5.4$ & \cite{Constantinescu2013} \\\hline
PBE0-D & 2.956 &  ~ &  10.91  &  44.01  & ~  & -2.62  & 63.08 & $\sim6.2$ & \cite{Constantinescu2013} \\\hline
PBE ($d=3.34$~\AA) & ~ &  ~ &  2.80  &  13.84  & ~  & 0.30  & 15.54  & $\sim2.0$ & \cite{Constantinescu2013} \\\hline
PBE0 ($d=3.34$~\AA) & ~ &  ~ &  4.14  &  15.80  & ~  & 0.74  & 17.38  & $\sim1.8$ & \cite{Constantinescu2013} \\\hline
PBE-D ($d=3.34$~\AA) & ~ &  ~ & 3.15 & 13.87 & ~  & 0.38  & 15.68 & $\sim1.9$ & \cite{Constantinescu2013} \\\hline
PBE0-D ($d=3.34$~\AA) & ~ &  ~ &  4.49  &  15.83 & ~  & 0.90  & 17.45 & $\sim1.8$ & \cite{Constantinescu2013} \\\hline
LDA & 3.245 &  ~ &  2  &  12 & ~  & 0  & 14 & ~ & \cite{Liu2003} \\\hline
LDA & 3.222 &  ~ &  1.6  &  11.7 & ~  & 0.6  & 12.1 & ~ & \cite{Ooi2006} \\\hline
\end{tabular}
}
\label{table:bn0_bulk}
\end{table*}

\begin{table*}
    \caption{Properties of hexagonal boron nitride bulk (for the layers aligned in the opposite directions: equilibrium interlayer distance $d_{\mathrm{eq}}$, shear modulus $C_{44}$, modulus for axial compression $C_{33}$, frequencies of relative in-plane and out-of-plane vibrations $f_E$ and $f_B$, respectively; for the layers aligned in the same direction: shear modulus $C'_{44}$ and frequency of relative in-plane vibrations $f'_E$) calculated using different functionals.}
   \renewcommand{\arraystretch}{1.2}
    \resizebox{\textwidth}{!}{
        \begin{tabular}{*{9}{c}}
\hline
Approach & $d_{\mathrm{eq}}$ (\AA) & $C_{44}$ (GPa) & $C_{33}$ (GPa) & $f_E$ (cm$^{-1}$) & $f_B$ (cm$^{-1}$) & $C'_{44}$ (GPa) & $f'_E$ (cm$^{-1}$) & Ref.  \\\hline
PBE-D2 &  3.088  &  8.62 &  63.40  & 64.58  & 175.17  & 9.08  & 66.29  & This work  \\\hline
PBE-D3 &  3.398  &  3.96 &  27.90 & 41.71  & 110.76  & 3.72  & 40.42  & This work  \\\hline
PBE-D3(BJ) &  3.318  &  4.78 & 28.15  & 46.40  & 112.59  & 4.92  & 47.09 & This work  \\\hline
PBE-TS &  3.317 &  2.98 &  37.81  & 36.65  & 130.51  & 4.50  & 45.04  & This work  \\\hline
PBE-TS/HI  &  3.386 &  ~ &  26.93  & ~ & 109.01 & ~  & ~  & This work  \\\hline
PBE-TS+SCS  &  3.326 &  2.93 &  36.45  & 36.27 & 127.99  & 3.71  & 40.82  & This work  \\\hline
optPBE-vdW &  3.388  &  3.91 &  29.71  & 41.52  & 114.47  & 3.76  & 40.74  & This work  \\\hline
vdW-DF2 &  3.478 &  3.58 &  31.36  & 39.21 & 116.08  & 3.12  & 36.60  & This work  \\\hline
PBE ($d=d_{\mathrm{exp}}$)  &  ~  &  5.21 &  ~  & 48.56 & ~  & 5.09 & 47.99  & This work  \\\hline
PBE-D2 ($d=d_{\mathrm{exp}}$) &  ~  &  5.09 &  ~  & 48.01  & ~  & 5.19  & 48.50 & This work  \\\hline
PBE-D3 ($d=d_{\mathrm{exp}}$) &  ~  &  5.47 &  ~  & 49.75  & ~  & 5.34  & 49.16 & This work  \\\hline
PBE-D3(BJ) ($d=d_{\mathrm{exp}}$) &  ~  &  5.04 &  ~  & 47.78  & ~  & 5.22  & 48.62  & This work  \\\hline
PBE-TS ($d=d_{\mathrm{exp}}$) &  ~  &  3.12 &  ~  & 37.56  & ~  & 4.66  & 45.94 & This work  \\\hline
PBE-TS+SCS ($d=d_{\mathrm{exp}}$) &  ~  &  3.07 &  ~  & 37.31  & ~  & 3.97 & 42.39 & This work  \\\hline
optPBE-vdW ($d=d_{\mathrm{exp}}$) &  ~  & 5.37 & ~  & 49.30  & ~  & 5.42  & 49.54  & This work  \\\hline
vdW-DF2 ($d=d_{\mathrm{exp}}$) &  ~  & 5.85 &  ~  & 51.47  & ~  & 5.51  & 49.97  & This work  \\\hline
LDA & 3.08 & ~ & ~ & ~ & 120 & ~ & ~ & \cite{Marini2006} \\\hline
LDA ($d=3.12$~\AA) & ~ & ~ & ~ & ~ & 90 & ~ & ~ & \cite{Marini2006} \\\hline
PBE ($d=3.12$~\AA) & ~ & ~ & ~ & ~ & 122 & ~ & ~ & \cite{Marini2006} \\\hline
EXX-RPA+ & 3.13 & ~ & ~ & ~ & $130\pm10$ & ~ & ~ & \cite{Marini2006} \\\hline
Exp. 10 K &  $3.3013\pm0.0010$  & ~ & ~ & ~ & ~ & ~ & ~ & \cite{Paszkowicz2002} \\\hline
Exp. 297.5 K & $3.3265\pm0.0010$ & ~ & ~ & ~ & ~ & ~ & ~ & \cite{Paszkowicz2002} \\\hline
Exp. 308.7 K &  $3.3300\pm0.0005$ & ~ & ~ & ~ & ~ & ~ & ~ & \cite{Pease1950} \\\hline
Exp. $308.2\pm0.5$ K &  $3.3306\pm0.0003$ & ~ & ~ & ~ & ~ & ~ & ~ & \cite{Pease1952} \\\hline
Exp. $\sim300$ K &  $3.330\pm0.004$ & ~ & ~ & ~ & ~ & ~ & ~ & \cite{Solozhenko1995,Solozhenko1997,Solozhenko2001} \\\hline
Exp. $\sim300$ K &  $3.341\pm0.001$ & ~ & ~ & ~ & ~ & ~ & ~ & \cite{Yoo1997} \\\hline 
Exp. $\sim300$ K &  $3.33\pm0.02$ & ~ & ~ & ~ & ~ & ~ & ~ & \cite{Fuchizaki2008} \\\hline
Exp. $\sim300$ K &  3.329 & $7.7\pm0.5$ & $27.0\pm0.5$ & ~ & ~ & ~ & ~ & \cite{Bosak2006} \\\hline
Exp. &  3.33 & ~ & ~ & ~ & ~ & ~ & ~ & \cite{Kobayashi2007, Kobayashi2008} \\\hline 
Exp. &  3.34--3.35 & $\le 3$ & ~ & ~ & ~ & ~ & ~ & \cite{Duclaux1992} \\\hline 
Exp. & ~ & ~ & 32 & ~ & ~ & ~ & ~ & \cite{Jager1977} \\\hline
Exp. &  ~ & ~ & ~ & 51.8 & ~ & ~ & ~ & \cite{Nemanich1981} \\\hline 
Exp. & ~ & ~ & ~ & ~ & 125 & ~ & ~ & \cite{Marini2006} \\\hline
Reference values & 3.301 & 5.4 & 29.5 & 51.8 & 125 & ~ & ~ & ~ \\\hline
\end{tabular}
}
\label{table:bn_bulk}
\end{table*}

\subsection{Interlayer distance and binding energy}
The interlayer distance is known with high precision for bulk materials \cite{Baskin1955,Zhao1989,Bosak2007, Trucano1975,Hanfland1989,Paszkowicz2002,Pease1950,Pease1952,Yoo1997,Fuchizaki2008,Bosak2006,Kobayashi2007, Kobayashi2008}  (Tables~\ref{table:graphite} and \ref{table:bn_bulk}). Since the energy calculations are performed at zero temperature, we consider as the reference for graphite the value of 3.336~\AA~corresponding to 4.2 K \cite{Baskin1955} and for boron nitride the value of 3.301~\AA~corresponding to 10 K \cite{Paszkowicz2002}. It should be noted, however, that increasing the temperature up to the room one does not lead to drastic changes in the interlayer distance and it grows only up to 3.36~\AA~for graphite \cite{Zhao1989,Bosak2007, Trucano1975,Hanfland1989} and to 3.33~\AA~for boron nitride \cite{Paszkowicz2002,Pease1950,Pease1952,Yoo1997,Fuchizaki2008,Bosak2006,Kobayashi2007, Kobayashi2008}. 

Very close to the room-temperature interlayer distance in graphite is the value of 3.35~\AA~obtained for bilayer and trilayer graphene on copper \cite{Brown2012} (Table~\ref{table:graphene}). Somewhat larger interlayer distances of 3.37--3.49~\AA~were measured for few-layer graphene on silicon carbide \cite{Weng2010, Varchon2007} and nickel \cite{Zhu2015}. These increased values can be explained by strong interaction of the corresponding substrates with graphene. To avoid consideration of such effects we use as the reference value for graphene bilayer the result from paper \cite{Brown2012}. The interlayer distances of $3.25\pm0.10$~\AA~\cite{Warner2010} and  $\sim3.5$~\AA~\cite{Zhi2009,Nag2010} measured for 10--20 layers of boron nitride are on the order of those for  bulk. However, the low accuracy of these data makes impossible their consideration as reference values.

All the considered exchange-correlation functionals  with account of vdW interactions give the equilibrium interlayer distances for graphene bilayer, graphite and bulk boron nitride within the relative deviation of 7\% from the reference values (Tables~\ref{table:graphene}, \ref{table:graphite} and \ref{table:bn_bulk}). Though this deviation is rather small, as discussed later, it is critical for prediction of properties of graphene and boron nitride layers associated with their relative motion. 

The experimental data on the binding energy in graph-ene-like materials are rather diverse, ranging from $-31$ meV/atom to $-52$ meV/atom \cite{Zacharia2004,Girifalco1956,Benedict1998,Liu2012} (note that all energies for bilayers in the present paper are given in meV per atom in the upper (adsorbed) layer). The QMC \cite{Mostaani2015,Spanu2009} and RPA approaches \cite{Lebegue2010} gave similar results in the range from $-36$ meV/atom to $-52$ meV/atom \cite{Mostaani2015,Spanu2009,Lebegue2010} (except for the recent RPA value of $-91$ meV/atom \cite{Zhou2015}). Therefore, as the reference value in the case we take the average of the experimental values. However, it should be kept in mind that the real binding energy lies within 40\% interval around this value. The considered functionals with account of vdW interactions give the binding energies for graphene bilayer and graphite in the upper limit of the experimental range or even beyond it (Tables~\ref{table:graphene} and \ref{table:graphite}). To our knowledge no experimental data on the binding energy in boron nitride is available.

\subsection{Potential energy surfaces}
The characteristics of the potential energy surfaces are difficult to access from the experimental studies directly. Thus, we do not introduce in this case any reference values. However, the potential energy surfaces are important for understanding of the properties associated with relative motion of the layers. In the following we discuss principal features of the potential energy surfaces of graphene and hexagonal boron nitride and compare our results with the previous calculations.  

\begin{figure*}
	\centering
	\includegraphics[width=\textwidth]{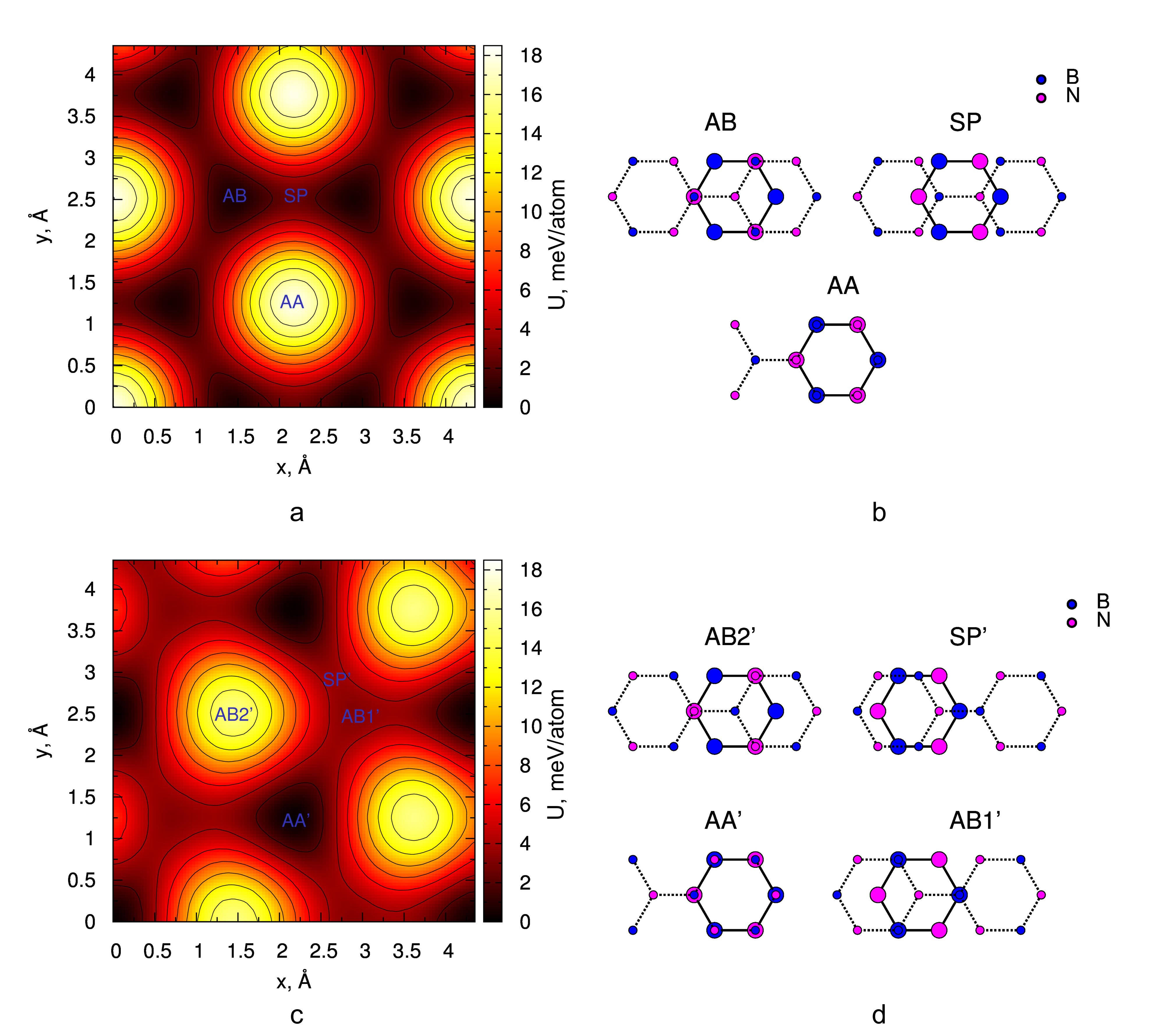}
	\caption{Interlayer interaction energy of hexagonal boron nitride bilayer $U$ (in meV per atom of the top layer) as a function of the relative displacement of the layers in the armchair ($x$, in \AA) and zigzag ($y$, in \AA) directions calculated using the vdW-DF2 functional at the interlayer distance of $d=$3.33~\AA: (a,b) boron nitride layers aligned in the same direction and (c,d) boron nitride layers aligned in the opposite directions. The energy is given relative to the AA' stacking. (b, d) Structures of the symmetric stackings are indicated. Boron and nitrogen atoms are coloured in blue and magenta, respectively.}
	\label{fig:pes}
\end{figure*}

In the case when boron nitride layers are aligned in the same direction the symmetry of the potential energy surface is the same as that of graphene (Fig.~\ref{fig:pes}a and b) \cite{Lebedev2016}. There are two equivalent energy minima AB in which half of the atoms of the adsorbed layer are located on top of the atoms of the bottom layer and the other half is on top of the hexagon centers. The saddle-point stacking SP that corresponds to the barrier to relative sliding of the layers is obtained by shifting the layers relative to each other by half of the bond length along the line connecting two adjacent minima AB. The maxima on the potential energy surface correspond to the AA stacking in which all atoms of the adsorbed layer are on top of the atoms of the bottom layer. 

The barriers for relative sliding of the layers in graphene bilayer and graphite $\Delta E_\mathrm{SP} = E_\mathrm{SP} - E_\mathrm{AB}$ calculated using different functionals range from 0.75~meV/atom to  2.35~meV/atom (Tables~\ref{table:graphene} and \ref{table:graphite}), in agreement with the previously reported values of 0.5 -- 2.1~meV/atom \cite{Kolmogorov2005, Aoki2007, Ershova2010, Lebedeva2011, Reguzzoni2012, Zhou2015}. The barriers obtained using the PBE-D2 and PBE-TS functionals are close to the estimates of the barrier from the experimental measurements of the shear mode frequency in bilayer, few-layer graphene and graphite of 1.7 meV/atom~\cite{Popov2012} and width of dislocations in few-layer graphene of 2.4 meV/atom~\cite{Alden2013}, respectively. The magnitude of corrugation of the potential energy surface $\Delta E_\mathrm{AA} = E_\mathrm{AA} - E_\mathrm{AB}$ varies from 7.3~meV/atom to 20.4~meV/atom according to different functionals, in agreement with the values of 6 -- 20~meV/atom obtained in previous DFT calculations \cite{Kolmogorov2005, Aoki2007, Ershova2010, Lebedeva2011, Reguzzoni2012, Zhou2015}, 8.8~meV/atom from the RPA calculations  \cite{Zhou2015} and $\sim12.4$ meV/atom from the QMC calculations \cite{Mostaani2015} (Tables~\ref{table:graphene} and \ref{table:graphite}). 

\begin{figure}
	\centering
	\includegraphics[width=0.5\textwidth]{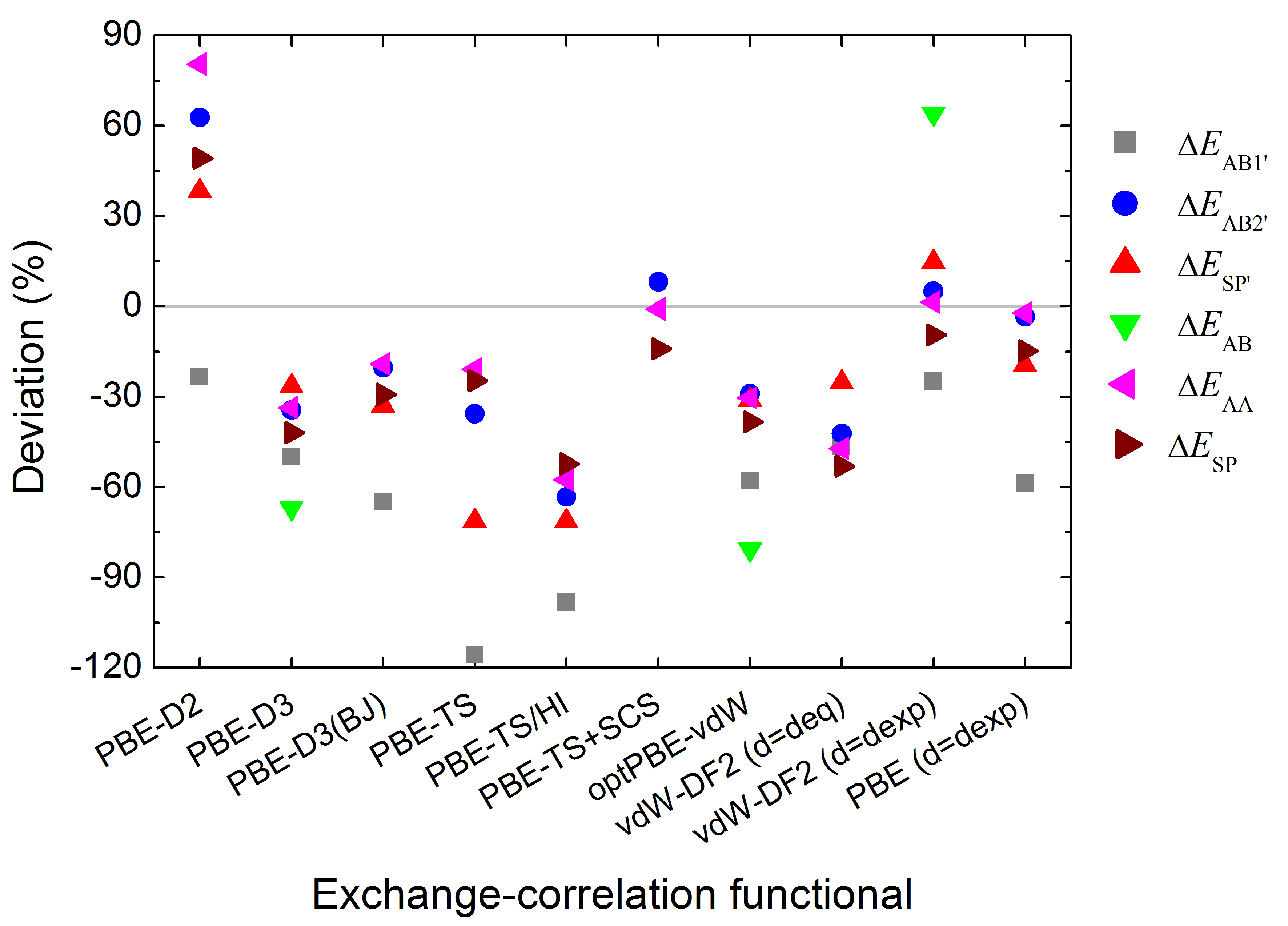}
	\caption{Relative deviation (in \%) of the characteristics of the potential energy surface of hexagonal boron nitride bilayer calculated using different vdW-corrected exchange-correlation functionals at the equilibrium interlayer distance from the LMP2 data \cite{Constantinescu2013}: (squares) $\Delta E_\mathrm{AB1'} = E_\mathrm{AB1'} - E_\mathrm{AA'}$, (circles) $\Delta E_\mathrm{AB2'} = E_\mathrm{AB2'} - E_\mathrm{AA'}$, (triangles up) $\Delta E_\mathrm{SP'} = E_\mathrm{SP'} - E_\mathrm{AA'}$, (triangles down) $\Delta E_\mathrm{AB} = E_\mathrm{AB} - E_\mathrm{AA'}$, (triangles left) $\Delta E_\mathrm{AA} = E_\mathrm{AA} - E_\mathrm{AB}$ and (triangles right) $\Delta E_\mathrm{SP} = E_\mathrm{SP} - E_\mathrm{AB}$. The results for the PBE functional without correction for vdW interactions are given for the experimental interlayer distance of $d_{\mathrm{exp}} = 3.30$~\AA. The results for the vdW-DF2 function are given both for the equilibrium ($d=d_{\mathrm{eq}}$) and experimental ($d=d_{\mathrm{exp}}$) interlayer distances. The data with deviation beyond -120\% and 90\% are not shown (see Table~\ref{table:bn0}). }
	\label{fig:dev_lmp2}
\end{figure}

The barriers $\Delta E_\mathrm{SP}$ for bilayer and bulk boron nitride with the layers aligned in the same direction are in general slightly larger than the results for graphene and graphite  (Tables~\ref{table:graphene} and \ref{table:graphite}) and for most of the considered functionals range from 1.1~meV/atom to 2.0~meV/atom (Tables~\ref{table:bn0} and \ref{table:bn0_bulk}). The only exception is the PBE-D2 functional that gives too large barriers of 3.6~meV/atom and 4.2~meV/atom for bilayer and bulk boron nitride, respectively. As discussed below, this is the consequence of the small equilibrium interlayer distance for this functional. The LMP2 value of $\sim2.4$~meV/atom \cite{Constantinescu2013} and RPA value of 2.1~meV/atom \cite{Zhou2015} are below the PBE-D2 result but higher than the results obtained other considered functionals  (Fig.~\ref{fig:dev_lmp2}). The calculated magnitudes of corrugation of the potential energy surface $\Delta E_\mathrm{AA}$ are 11 -- 20~meV/atom for most of the functionals and 35 -- 40~meV/atom for PBE-D2 (Tables~\ref{table:bn0} and \ref{table:bn0_bulk}). These values agree with the previously reported DFT data for boron nitride of 7 -- 60~meV/atom \cite{Liu2003,Ooi2006, Constantinescu2013, Zhou2015} as well as the LMP2  \cite{Constantinescu2013} (Fig.~\ref{fig:dev_lmp2}) and RPA \cite{Zhou2015} results of about 20 meV/atom and 11.4 meV/atom, respectively.

In the case of boron nitride with the layers aligned in the opposite directions, the potential energy surface has two inequivalent energy minima (Fig.~\ref{fig:pes}c and d). The first energy minimum corresponds to the AA' stacking with all atoms of the adsorbed layer in the on-top position. The second energy minimum corresponds to the AB1' stacking in which the boron atoms of the adsorbed layer are on top of the boron atoms of the bottom layer and the nitrogen atoms are on top of the hexagon centers. The AB2' stacking with the nitrogen atoms in the on-top position and the boron atoms are on top of the hexagon centers has the maximal energy among all boron nitride stackings for the layers aligned in the opposite directions. The saddle-point stacking SP' corresponding to the barrier to relative sliding of the layers aligned in the opposite directions can be found by shifting the layers along the straight line corresponding to the transition between the AB1' and AA' stackings.

Almost all the considered functionals agree that the energies of the AA', AB1' and AB stackings of boron nitride layers are rather close and differ only by several meV/atom (Tables~\ref{table:bn0} and \ref{table:bn0_bulk}). However, the order of stability of these structures varies depending on the approach, which is supported by the literature data \cite{Liu2003,Ooi2006,Nag2010, Marom2010,Constantinescu2013,Zhou2015}. While the experimental studies \cite{Pease1950}, RPA \cite{Zhou2015} and LMP2 \cite{Constantinescu2013} results and DFT calculations with account of nonlocal many-body dispersion (MBD) \cite{Gao2015} suggest that the ground state of boron nitride corresponds to the AA' stacking, the most favourable stacking according to the PBE-D2 and PBE-D3(BJ) functionals is AB. The PBE-TS, PBE-TS/HI  and PBE-TS+SCS functionals predict that  both the AB1' and AB stackings are lower in energy than the AA stacking in boron nitride bulk (Table~\ref{table:bn0_bulk}). For boron nitride  bilayer, the PBE-TS functional gives the similar result, the PBE-TS/HI  shows the preference of the AB stacking over the AA' stacking, while according to the PBE-TS+SCS functional, the SP' and AB1' stackings  stran-gely coincide and have the clearly overestimated relative energy of $\Delta E_\mathrm{AB1'} = E_\mathrm{AB1'} - E_\mathrm{AA'}=$ 8.6 meV/atom (Table~\ref{table:bn0}). 

The only considered functionals that correctly describe the ground state of boron nitride are PBE-D3 and vdW-DF2. According to them, the energy difference between the AB and AA' stackings $\Delta E_\mathrm{AB} = E_\mathrm{AB} - E_\mathrm{AA'}$ is within 0.6 meV/atom, in agreement with the LMP2 \cite{Constantinescu2013} and RPA \cite{Zhou2015} results and the experimental data \cite{Warner2010} that both the AA' and AB1' stackings can be observed for  boron nitride bilayer. It should be nevertheless noted that the optPBE-vdW functional gives the correct order of stackings in energy for boron nitride bilayer, while for bulk the AB stacking is preferred over the AA' stacking only by 0.02 meV/atom, i.e. within the typical error of DFT calculations.

The barrier for transition from the AA' stacking to the AB1' one $\Delta E_\mathrm{SP'} = E_\mathrm{SP'} - E_\mathrm{AA'}$ varies for the PBE-D3, PBE-D3(BJ), optPBE-vdW and vdW-DF2 functionals from 2.3~meV/atom to 2.9~meV/atom  (Tables~\ref{table:bn0} and \ref{table:bn0_bulk}). The PBE-D2 functional gives remarkably higher values of 4.7 meV/atom and 5.6 meV/atom for boron nitride bilayer and bulk, respectively, while the PBE-TS and PBE-TS/HI  functionals predict the smaller values in the 0.6 -- 1.0 meV/atom range. The PBE-TS+SCS  functional provides the similarly small value of 0.7 meV/atom for the bulk material, while the result for bilayer of 8.6 meV/atom is clearly too large. For comparison, the LMP2 method gives the barrier for boron nitride bilayer of 3.4 meV/atom \cite{Constantinescu2013} (Fig.~\ref{fig:dev_lmp2}). The magnitudes of corrugation of the potential energy surface for boron nitride with the layers aligned in the opposite directions $\Delta E_\mathrm{AB2'} = E_\mathrm{AB2'} - E_\mathrm{AA'}$ range from 10~meV/atom to 30~meV/atom, similar to the values 7 -- 44 meV/atom reported in literature for different DFT methods \cite{Liu2003,Ooi2006, Constantinescu2013,Zhou2015} and including the LMP2 \cite{Constantinescu2013} and RPA \cite{Zhou2015} values of about 16 meV/atom and 11~meV/atom, respectively.

Clearly overestimated or underestimated corrugations of the potential energy surfaces obtained using different functionals can be attributed to too small or too large equilibrium interlayer distances. It is known from previous publications \cite{Lebedeva2011, Reguzzoni2012} that the corrugations of the potential energy surface increase exponentially upon decreasing the interlayer distance $d$. Using the experimental interlayer distance of $d_{\mathrm{exp}}=3.34$~\AA~for graphene and graphite and $d_{\mathrm{exp}}=3.30$~\AA~for boron nitride, provides more reasonable values for characteristics of the potential energy surface for all the considered approches except the DFT-TS family. 

For graphene and graphite, the barrier $\Delta E_\mathrm{SP}$ and magnitude of corrugation $\Delta E_\mathrm{AA}$  obtained for the experimental interlayer distance using the PBE-D2, PBE-D3, PBE-D3(BJ), optPBE-vdW and vdW-DF2 functionals lie in the narrow ranges from 1.55 meV/atom to 1.65 meV/atom and from 13.7 meV/atom to 14.9 meV/atom, respectively (Tables~\ref{table:graphene} and \ref{table:graphite}). Therefore, using the equilibrium interlayer distance, the agreement with the estimate of the barrier from the experimental measurements of the shear mode frequency in bilayer, few-layer graphene and graphite of 1.7 meV/atom~\cite{Popov2012} is improved compared to the results for the equilibrium interlayer distance. 

For boron nitride with the layers aligned in the same direction, the barrier $\Delta E_\mathrm{SP}$ and magnitude of corrugation $\Delta E_\mathrm{AA}$ obtained for the experimental interlayer distance using the PBE-D2, PBE-D3, PBE-D3(BJ), optPBE-vdW and vdW-DF2 functionals are 2.03 -- 2.25 meV/atom and 19.0 -- 20.0 meV/atom (Tables~\ref{table:bn0} and \ref{table:bn0_bulk}), respectively, in better agreement with the LMP2 values of $\sim2.4$~meV/atom and $\sim20$ meV/atom \cite{Constantinescu2013} (Fig.~\ref{fig:dev_lmp2}). For boron nitride with the layers aligned in the opposite directions at the experimental interlayer distance, the characteristics $\Delta E_\mathrm{AB2'}$ and $\Delta E_\mathrm{SP'}$ are 15.9 -- 17.5 meV/atom and 2.67 -- 3.90 meV/atom, respectively. These values are also closer to the LMP2 data of 16.1 -- 16.5 meV/atom and 3.4 meV/atom \cite{Constantinescu2013} (Fig.~\ref{fig:dev_lmp2}). 

Therefore, the use of the experimental interlayer distance for most of the considered functionals improves the description of the potential energy surfaces both for graphite and boron nitride. This, however, does not solve the problem of the wrong ground-state stacking of boron nitride for the PBE-D2, PBE-D3(BJ) and optPBE-vdW functionals (Tables~\ref{table:bn0} and \ref{table:bn0_bulk}). In the case of the PBE-D3 functional the energy of the AB stacking of  boron nitride bilayer relative to the AA' stacking also becomes slightly negative at the experimental interlayer distance. The functionals different from PBE, such as rPBE \cite{rpbe}, revPBE \cite{revpbe} and PBEsol \cite{pbesol}, in the pure form or with vdW corrections according to the DFT-D3 or DFT-D3(BJ) methods, suffer from the same deficiency. The use of the hybrid functionals, such as PBE0 \cite{pbe0}, can be a solution (Table~\ref{table:bn0}, these calculations were performed using the $16\times 24\times 1$ k-point grid and the maximum kinetic energy of plane waves of 500 eV). Nevertheless, such functionals are heavy computationally. Among the considered standard functionals, vdW-DF2 is the only one that describes the order of the metastable states of boron nitride at the experimental interlayer distance both for bulk and bilayer. Moreover, using this approach, the relative energies of the AA', AB and AB1' stackings of hexagonal boron nitride are not just qualitatively correct but also quantitatively close to the results of the LMP2 calculations \cite{Constantinescu2013}  (Tables~\ref{table:bn0} and \ref{table:bn0_bulk}, Fig.~\ref{fig:dev_lmp2}). 

It should also be noted that the characteristics of the potential energy surfaces obtained at the experimental interlayer distances using the PBE-D2, PBE-D3, PBE-D3(BJ), optPBE-vdW and vdW-DF2 functionals are very close to those calculated using the PBE functional without any vdW correction (Tables~\ref{table:graphene}--\ref{table:bn0} and \ref{table:bn0_bulk}). This is consistent with previous studies \cite{Ershova2010, Lebedeva2011,
Reguzzoni2012,Marom2010,Constantinescu2013, Lebedev2016}, which suggest that the contribution of van der Waals interaction into the corrugation of the potential energy surface at a given interlayer distance is negligibly small. Similar observations hold also for the rPBE, revPBE and PBEsol functionals (Tables~\ref{table:bn0} and \ref{table:bn0_bulk}).

Let us now consider the difference in the characteristics of the potential energy surfaces $\Delta E_{\mathrm{AB2'}}$, $\Delta E_{\mathrm{SP'}}$, $\Delta E_{\mathrm{AA}}$ and $\Delta E_{\mathrm{SP}}$ for bilayer  (Tables~\ref{table:graphene} and \ref{table:bn0})  and bulk   (Tables~\ref{table:graphene} and \ref{table:bn0_bulk})  materials. This difference
can be associated with the changes in the equilibrium interlayer distance and interaction of non-adjacent layers in the bulk. Taking into account variation of the equilibrium interlayer distance, the differences in the listed properties for the PBE-D2, PBE-D3, PBE-D3(BJ), vdW-DF2 and optPBE-vdW functionals are within 20\%. At the experimental interlayer distance only the effect of interaction of non-adjacent layers in the bulk is left and the differences in the characteristics of the potential energy surfaces are within 5\%. Therefore, for these approaches, the effect of interaction of non-adjacent layers in bulk on the characteristics of the potential energy surface is much smaller compared to the changes in these characteristics due to variation of the equilibrium interlayer distance. 

The much more prominent effect of non-adjacent layers is observed for the DFT-TS family of methods, especially for the DFT-TS+SCS approach, where the potential energy surfaces of boron nitride with the layers aligned in the opposite directions are qualitatively different for the bulk and bilayer materials. As mentioned above, according to this method, the SP' and AB1' stackings of the bilayer coincide and have a very high energy relative to the AA' stacking. Such a remarkable modification of the potential energy surface should be attributed to perculiarities of the screening effects in two-dimensional materials.

\subsection{Frequencies and elastic properties} 
While the barrier to relative sliding of the layers and magnitude of the corrugation of the potential energy surfaces are difficult to measure directly, some information on the potential energy surfaces is contained in the available experimental data on the shear modulus $C_{44}$ \cite{Bosak2007, Nicklow1972,Grimsditch1983,Seldin1970,Tan2012,Bosak2006,Duclaux1992} and frequency $f_E$ of the shear mode $E_{2g}$ \cite{Mohr2007, Brillson1971,Tan2012, Boschetto2013,Hanfland1989,Nicklow1972,Boschetto2013,Nemanich1981}. The information describing the change of the binding energy with the interlayer distance corresponds to the modulus $C_{33}$ for axial compression \cite{Bosak2007,Nicklow1972,Blakslee1970,Gauster1974,Wada1980,Alzyab1988,Boschetto2013, Tan2012,Bosak2006,Jager1977} and frequency $f_B$ of the out-of-plane $B_{1g}$ mode \cite{Lui2012,Mohr2007, Brillson1971, Alzyab1988,Nicklow1972,Lui2014,Marini2006}. As the reference data in these cases we use the average of the experimental values using that normally the experimental data fall within narrow intervals around the average values (Tables~\ref{table:graphite}, \ref{table:graphene} and \ref{table:bn_bulk}). The error in the experimental measurements related to scatter of the data for graphite is within 9\% and 13\% for the moduli $C_{33}$ \cite{Bosak2007,Nicklow1972,Blakslee1970,Gauster1974,Wada1980,Alzyab1988} and $C_{44}$ \cite{Bosak2007, Nicklow1972,Grimsditch1983,Seldin1970,Tan2012}, respectively, and within 1\% and 3\% for the frequencies $f_B$ \cite{Lui2012,Mohr2007, Brillson1971, Alzyab1988,Nicklow1972} and $f_E$ \cite{Mohr2007, Brillson1971,Tan2012, Boschetto2013,Hanfland1989,Nicklow1972}, respectively. The errors about 3\% and 10\% are associated with measurements of the frequencies $f_B$ \cite{Lui2012, Lui2014} and $f_E$ \cite{Boschetto2013, Tan2012} in graphene bilayer. For boron nitride, the experimental data are available only for bulk \cite{Bosak2006,Duclaux1992,Jager1977,Nemanich1981,Marini2006}. The errors related to scatter of the data for the moduli $C_{44}$ \cite{Bosak2006,Duclaux1992} and $C_{33}$ \cite{Bosak2006,Jager1977} are estimated as 40\% and 8\%, respectively.

In our calculations, the shear mode frequency $f_E$ of bilayer in which the layers slide rigidly in the opposite in-plane directions is determined from the curvature $\partial^2 U/\partial x^2$ of the potential energy surface for in-plane displacements out of a given energy minimum~\cite{Popov2012}
\begin{equation} \label{eq_freq}
	\begin{split}
		f_E = \frac{1}{2\pi}\sqrt{\frac{1}{\mu}\frac{\partial^2 U}{\partial x^2}},
	\end{split}
\end{equation}
where $\mu = m/2$ and $m$ is the average mass of atoms in the layers. In bulk materials, the frequency is additionally multiplied by the factor of $\sqrt{2}$ \cite{Popov2012}. To find the curvature of the potential energy surface in-plane displacements up to 0.05~\AA~are considered along the armchair direction and the obtained energy curve is approximated by a parabola. For boron nitride, the calculations are performed for the AA' and AB stackings, which are very close in energy and can be both observed for bilayer \cite{Warner2010}. Since the PBE-TS/HI  functional gives the non-parabolic dependences of potential energy on in-plane displacement around the AA' and AB stackings, we do not consider the properties characterizing relative in-plane motion of the layers in this case.

The same curvature $\partial^2 U/\partial x^2$ of the potential energy surface also determines the shear modulus $C_{44}$, which has the same formula for bilayer and bulk
\begin{equation} \label{eq_modulus}
	\begin{split}
		C_{44} = \frac{d}{\sigma}\frac{\partial^2 U}{\partial x^2},
	\end{split}
\end{equation}
where $\sigma =3\sqrt{3}l^2/4$ is the area per atom ($l$ is the bond length) and $d$ is the interlayer
distance. 

The frequency $f_B$ of relative out-of-plane vibrations of the layers and modulus for axial compression $C_{33}$ are found by equations similar to eqs.~\ref{eq_freq} and \ref{eq_modulus} in which the curvature of the potential energy surface is taken for out-of-plane displacements $\partial^2 U/\partial  z^2$. Out-of-plane displacements up to 0.5\% of the equilibrium interlayer distance are considered and the obtained energy curve is approximated by a parabola. The accuracy of description of the elastic moduli and frequencies by different functionals is analyzed in detail in subsection 3.4. 

\subsection{Discussion}
\begin{figure}
	\centering
	\includegraphics[width=0.5\textwidth]{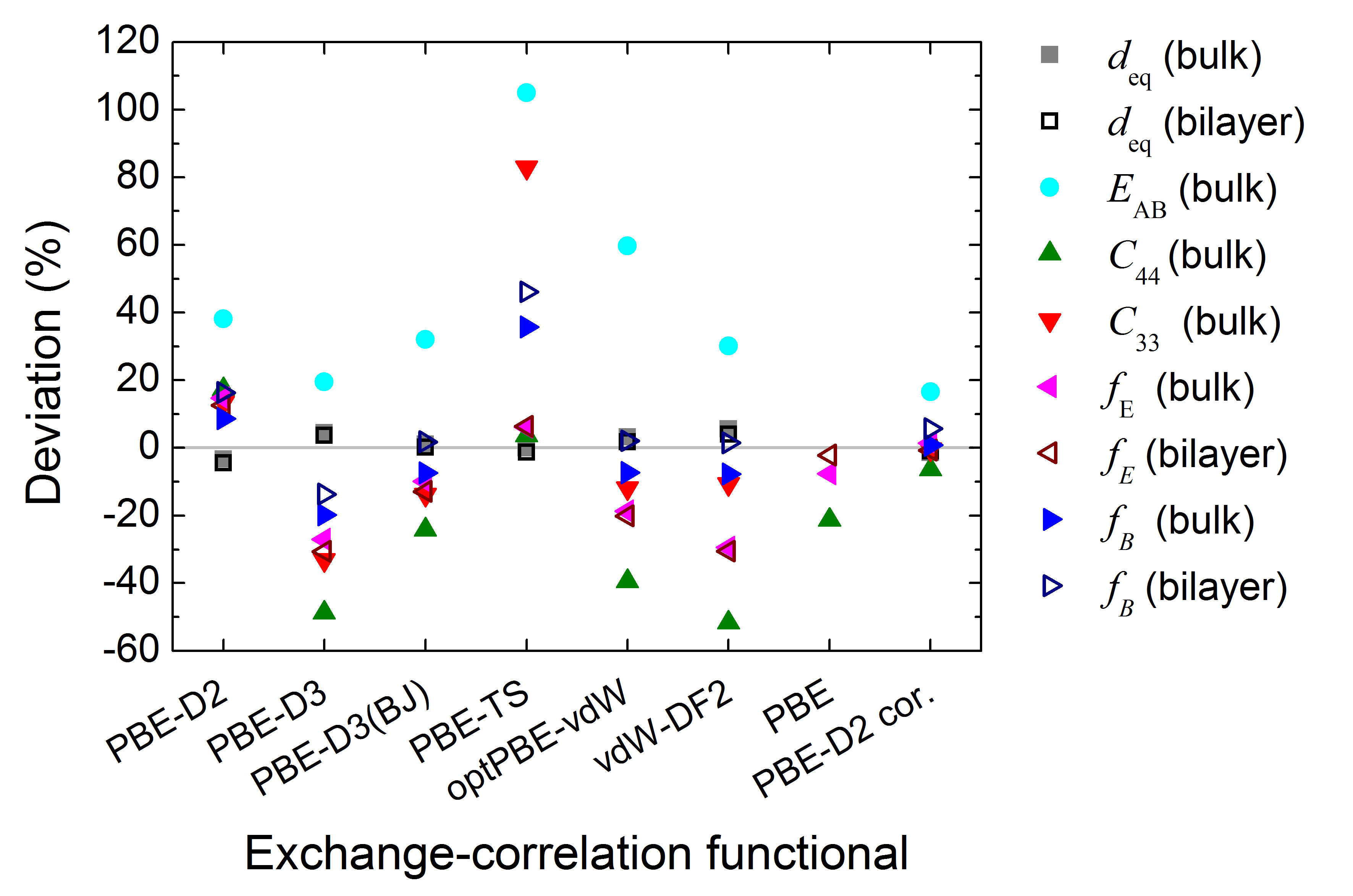}
	\caption{Relative deviation (in \%) of the properties of graphite (filled symbols) and graphene bilayer (open symbols) calculated using different vdW-corrected exchange-correlation functionals from the reference experimental data: (squares) equilibrium interlayer distance $d_{\mathrm{eq}}$, (circles) binding energy in the AB stacking $E_{\mathrm{AB}}$, (triangles up) shear modulus $C_{44}$, (triangles down) modulus for axial compression $C_{33}$, (triangles left) shear mode frequency $f_E$ and (triangles right) frequency of relative out-of-plane vibrations $f_B$. The results for the PBE functional without correction for vdW interactions are given for the experimental interlayer distance of $d_{\mathrm{exp}} = 3.34$~\AA.}
	\label{fig:dev_graph}
\end{figure}

Let us first discuss the results of calculations in connection with the available experimental data for graphene bilayer and graphite (Tables~\ref{table:graphene} and \ref{table:graphite}, respectively). Deviations of the physical properties of graphene bilayer and graphite from the reference data obtained by averaging the experimental values are shown in Fig.~\ref{fig:dev_graph}. It is seen that none of the functionals considered is able to reproduce these data fully. Similar descrepancies in the equilibrium lattice parameters, cohesive energy, interlayer binding energy, bulk modulus, $C_{33}$ and $C_{333}$ moduli and $\pi_z$-splitting of graphite were observed previously in Ref. \cite{Rego2016}.

The closest agreement with the reference data is achiev-ed for the PBE-D2 approach. The functional gives the magnitude of the binding energy $E_{\mathrm{AB}}$ of graphite in the AB stacking at the upper limit of the experimental range (Table~\ref{table:graphite}, 40\% deviation from the reference value) and overestimates the shear modulus $C_{44}$, modulus $C_{33}$ for axial compression and frequencies of relative in-plane $f_E$ and out-of-plane $f_B$ vibrations in  graphene bilayer and graphite by 8--18\%. The exaggeration of the characteristics related to in-plane displacement of the layers can be partially attributed to underestimation of the equilibrium interlayer distance $d_{\mathrm{eq}}$ by 3.4\% and 2.8\% in graphite and graphene bilayer, respectively. Though this deviation does not look large, the corrugations of the potential energy surface increase exponentially upon decreasing the interlayer distance $d$ \cite{Lebedeva2011, Reguzzoni2012}. Therefore, even a small deviation in the equilibrium interlayer distance $d_{\mathrm{eq}}$ results in a considerable error in the properties related to in-plane motion of the layers, such as the shear modulus $C_{44}$ and shear mode frequency $f_E$. The underestimation of the equilibrium interlayer distance $d_{\mathrm{eq}}$ combined with overestimation of the binding energy $E_{\mathrm{AB}}$ can be responsible for overestimation of the properties related to out-of-plane motion of the layers including the modulus $C_{33}$ for axial compression and the frequency $f_B$ of out-of-plane vibrations.

Similar in magnitude deviations from the reference values are observed for the PBE-D3(BJ) functional (Fig.~\ref{fig:dev_graph}). The main difference from the PBE-D2 approach, however, is that the elastic moduli $C_{44}$  and $C_{33}$ and frequencies $f_E$ and $f_B$ are underestimated not overestimated. Such an underestimation for the properties $C_{44}$ and $f_E$ related to in-plane motion of the layers correlates with the overestimation of the equilibrium interlayer distance $d_{\mathrm{eq}}$ by 1.1\% and 1.8\% in graphite and graphene bilayer, respectively.

The PBE-TS functional reproduces very well the equilibrium interlayer distance $d_{\mathrm{eq}}$ (within 0.3\%, Fig.~\ref{fig:dev_graph}) and the properties associated with in-plane relative displacement of the layers, the shear modulus $C_{44}$ and shear mode frequency $f_E$, both for graphite and graphene bilayer (with-in 6\%). However, it strongly overestimates the binding energy $E_{\mathrm{AB}}$ in graphite (by 105\%), frequency $f_B$ of relative out-of-plane vibrations and modulus $C_{33}$ for axial compression (36\% and 83\% for graphite, respectively). 

The PBE-D3, vdW-DF2 and optPBE-vdW functionals noticeably overestimate the equilibrium interlayer distance $d_{\mathrm{eq}}$ (by 3--6\%, Fig.~\ref{fig:dev_graph}). Correspondingly the shear mode frequencies $f_E$ and shear modulus $C_{44}$ are strongly underestimated (by 20--30\% and 40--50\%, respectively). Though the binding energy $E_{\mathrm{AB}}$ in graphite is within the experimental range (PBE-D3, vdW-DF2) or overestimated (optPBE-vdW), the frequency $f_B$ of out-of-plane vibrations and modulus $C_{33}$ for axial compression in graphite are underestimated (by 7--20\% and 10--33\%, respectively).

\begin{figure}
	\centering
	\includegraphics[width=0.5\textwidth]{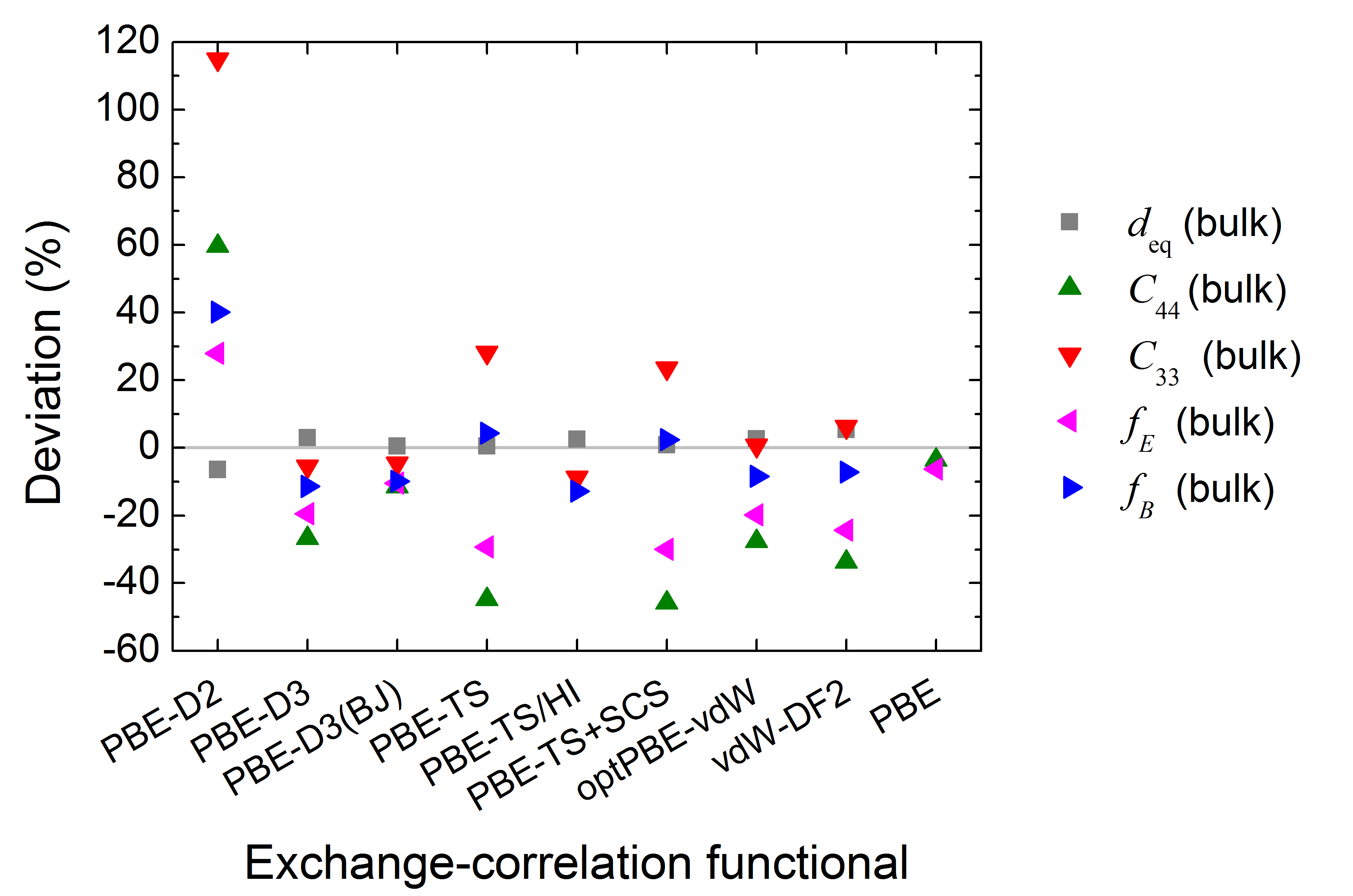}
	\caption{Relative deviation (in \%) of the properties of hexagonal boron nitride bulk calculated using different vdW-corrected exchange-correlation functionals from the reference experimental data: (squares) equilibrium interlayer distance $d_{\mathrm{eq}}$, (triangles up) shear modulus $C_{44}$, (triangles down) modulus for axial compression $C_{33}$, (triangles left) shear mode frequency $f_E$ and (triangles right) frequency of relative out-of-plane vibrations $f_B$. The results for the PBE functional without correction for vdW interactions are given for the experimental interlayer distance of $d_{\mathrm{exp}} = 3.30$~\AA.}
	\label{fig:dev_bn}
\end{figure}

It is, however, premature to judge about the accuracy of different approaches on the basis of the results  only for graphene bilayer and graphite. Indeed the comparison of the calculation results obtained using the PBE-D2 and PBE-TS functionals with the experimental data for hexagonal boron nitride bulk (Fig.~\ref{fig:dev_bn}) gives a picture very different from that for graphite. The equilibrium interlayer distance is much more underestimated by the PBE-D2 approach in boron nitride bulk (6.5\%) compared to graphite (3.4\%), which results in strongly overestimated properties related both to in-plane and out-of-plane relative motion of boron nitride layers (30--120\%). The PBE-TS functional describes better the frequency $f_B$ of relative out-of-plane vibrations and modulus $C_{33}$ for axial compression (4.4\% and 28\% overestimation, respectively) compared to graphite but fails to reproduce the properties of boron nitride related to in-plane motion of the layers, the shear modulus $C_{44}$ and shear mode frequency $f_E$ (45\% and 29\% underestimation, respectively). Similar errors are observed in the properties characterizing relative in-plane motion of the layers for the PBE-TS+SCS functional, while evaluation of these quantities using the PBE-TS/HI  functional is complicated by the non-parabolic dependences of interlayer interaction energy on displacement around the AA' and AB stackings.

The PBE-D3, PBE-D3(BJ), optPBE-vdW and vdW-DF2 functionals give more consistent results in the calculations for graphite and boron nitride (Fig.~\ref{fig:dev_bn}). There is even some improvement for boron nitride. The closer approximation of the interlayer distance by these approaches in boron nitride relative to graphite is accompanied by the improved accuracy in the frequencies and moduli related both to in-plane and out-of-plane relative motion of the layers.

Based on the data both for graphite and boron nitride it can be concluded that the best description of the properties of the AB and AA' stackings, respectively, which are known to be the ground-state ones from the experimental studies, is provided by the PBE-D3(BJ) functional (Figs.~\ref{fig:dev_graph} and \ref{fig:dev_bn}). However, as discussed in subsection 3.2, this approach along with PBE-D2, PBE-TS, PBE-TS/HI, PBE-TS+SCS and optPBE-vdW fails to predict the correct order in energy for the AA', AB1' and AB stackings of boron nitride. The functionals that adequately describe the relative stability of these metastable stackings, PBE-D3 and vdW-DF2, are comparable in accuracy (Figs.~\ref{fig:dev_graph} and \ref{fig:dev_bn}). The PBE-D3 functional is somewhat more precise in the properties associated with in-plane motion of the layers, the shear mode frequency  $f_E$  and shear modulus $C_{44}$, while the vdW-DF2 approach gives better results for the properties associated with out-of-plane motion of the layers, the frequency $f_B$ of relative out-of-plane vibrations and modulus $C_{33}$ for axial compression.

Considering only the latter group of properties, the results closest to the reference experimental data are obtained for three functionals: PBE-D3(BJ), optPBE-vdW and vdW-DF2. However, the use of the vdW-DF2 functional is preferred considering the correct ordering of the metastable states of hexagonal boron nitride in energy.

\begin{figure}
	\centering
	\includegraphics[width=0.5\textwidth]{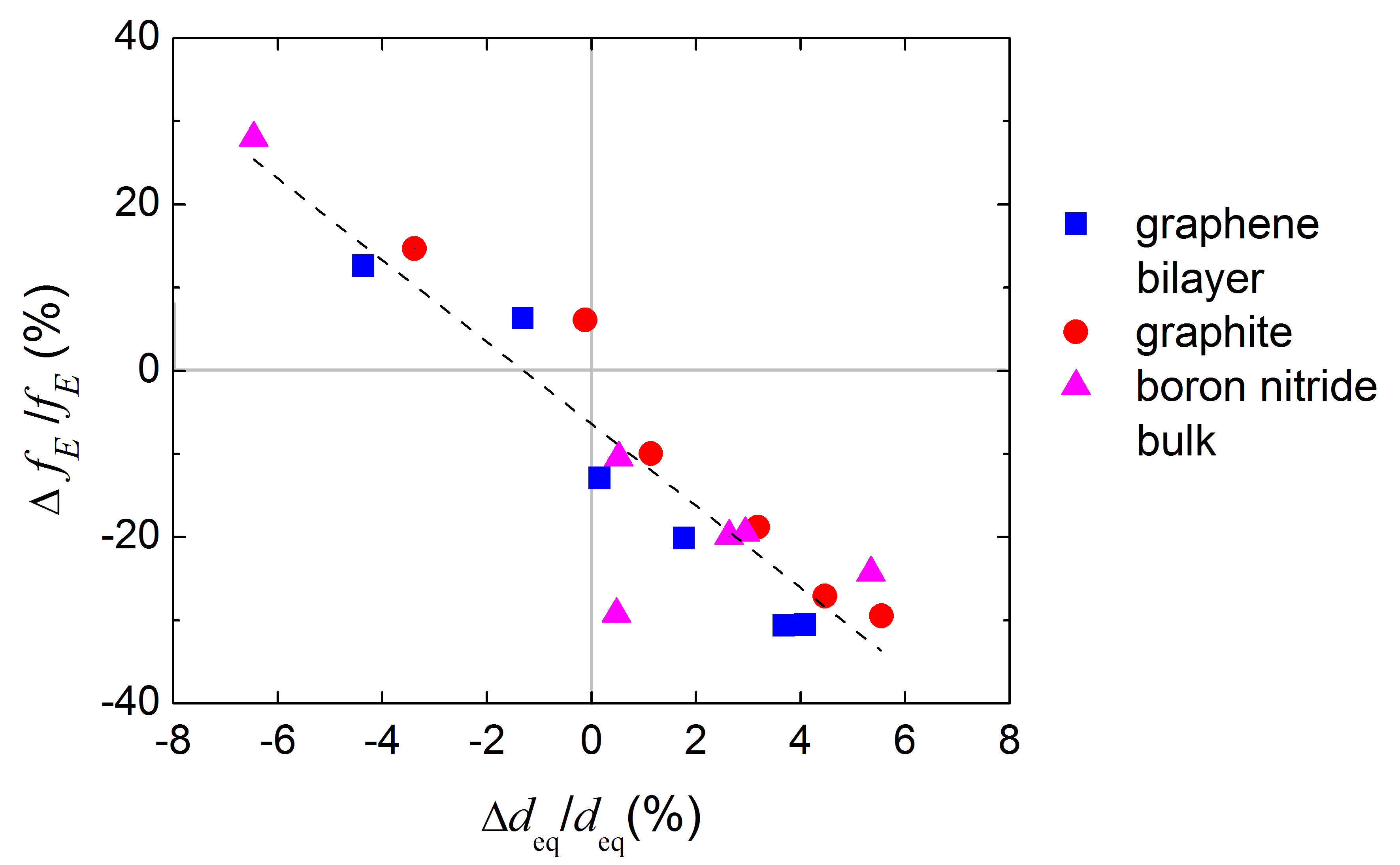}
	\caption{Calculated relative deviation  $\Delta f_E/f_E$ of the shear mode frequency (in \%) as a function of the relative deviation $\Delta d_{\mathrm{eq}}/d_{\mathrm{eq}}$ of the equilibrium interlayer distance (in \%) for PBE-D2, PBE-D3, PBE-D3(BJ), PBE-TS, vdW-DF2 and optPBE-vdW functionals: (squares) graphene bilayer, (circles) graphite and (triangles) hexagonal boron nitride bulk. The dashed line corresponding to the linear approximation of the dependence is shown to guide the eye.}
	\label{fig:dev_freq}
\end{figure}

The comparison of the results obtained using the vdW-corrected functionals for graphene bilayer, graphite and boron nitride (Figs.~\ref{fig:dev_graph} and \ref{fig:dev_bn}) with the reference data suggests that the approaches that guess better the equilibrium interlayer distance are more precise for the properties associated with in-plane motion of the layers (Fig.~\ref{fig:dev_freq}). In fact, using the experimental interlayer distance for the PBE-D2, PBE-D3, PBE-D3(BJ), vdW-DF2 and optPBE-vdW functionals improves not only description of the characteristics of the potential energy surface, as discussed in section 3.2, but also of the shear moduli and shear mode frequencies (Tables~\ref{table:graphene} -- \ref{table:bn_bulk}). 

The data obtained using these methods at the experimental interlayer distance are also very close to the results for the pure PBE functional without corrections for vdW interactions (Tables~\ref{table:graphene} -- \ref{table:bn_bulk}). For the latter, the shear mode frequency $f_E$ and shear modulus $C_{44}$ of graphite calculated at the experimental interlayer distance are 7.7\% and 21\% below the reference data, respectively (Fig.~\ref{fig:dev_graph}). For  graphene bilayer, the calculated shear mode frequency $f_E$ is within the experimental range. For boron nitride, the shear mode frequency $f_E$ and shear modulus $C_{44}$ calculated at the experimental interlayer distance are 3.5\% and 6.3\% below the reference data, respectively (Fig.~\ref{fig:dev_bn}). Therefore, the use of the experimental interlayer distance seems to be appropriate for description of any properties related to in-plane relative displacement of the layers. 

Again at the experimental interlayer distance most of the functionals including the pure PBE functional fail to describe the order of the metastable states of boron nitride in energy (Tables~\ref{table:bn0} and \ref{table:bn0_bulk}). Among the standard, i.e. non-hybrid, functionals, the correct relative energies of the metastable states of boron nitride both in the forms of bulk and bilayer are predicted in this case only by the vdW-DF2 functional. Thus this is the most adequate functional for calculations at the experimental interlayer distance.

It should be noted that agreement of the methods from the DFT-D family with the reference experimental data can be improved by readjustment of their semi-empirical parameters for specific materials. For example, the use of the dispersion coefficient $C_6^\mathrm{C}= 1.1725$ $\mathrm{J}\cdot \mathrm{nm}^6/\mathrm{mol}$ and van der Waals radius $R_0^\mathrm{C}=1.4868$~\AA~for carbon in the DFT-D2 approach (with $s_6 = 1$) leads to much better description of the properties of graphene bilayer and graphite (see the results for the PBE-D2 corrected functional in Tables~\ref{table:graphene} and \ref{table:graphite}). In this case the relative deviation of the binding energy $E_{\mathrm{AB}}$ of graphite from the reference value is reduced to 16.5\%, while the moduli $C_{44}$ and $C_{33}$ and frequencies $f_E$ and $f_B$ show the deviations within 6.5\% (Fig.~\ref{fig:dev_graph}).

\section{Conclusions}
The DFT calculations of physical properties of bilayer graphene, graphite, bulk and bilayer boron nitride have been performed using different exchange-correlation functionals including corrections for vdW interactions. The performance of the functionals with respect to such experimentally measurable quantities as the equilibrium interlayer distance, binding energy, frequencies of relative in-plane and out-of-plane vibrations of the layers, shear modulus and modulus for axial compression has been compared.

It is shown that the best description of the listed properties for the AB stacking of  graphene bilayer and graphite and AA' stacking of boron nitride, which are known to be the ground-state ones from the experimental studies, is provided by the PBE-D3(BJ) functional. However, this approach fails to predict the order of the metastable states of boron nitride in energy. With account of this ordering, the best options are the PBE-D3 and vdW-DF2 functionals. 

The properties associated with relative out-of-plane motion of the layers, the frequency of relative out-of-plane vibrations and modulus for axial compression, are especially relevant under external load. For these characteristics, the results of the PBE-D3(BJ), optPBE-vdW and vdW-DF2 functionals are the ones closest to the corresponding reference experimental data. The use of the vdW-DF2 functional is again preferred considering the correct ordering of the metastable states of boron nitride in energy. 

Significant improvement in the properties related to in-plane relative motion of the layers, shear mode frequency and shear modulus, is achieved when the experimental equilibrium interlayer distance is used. In this case it is also recommended to use the vdW-DF2 functional, which adequately describes energetics of the metastable states of boron nitride. In addition, the use of the experimental equilibrium interlayer distance improves description of the characteristics of the potential energy surfaces, such as the barriers to relative sliding of the layers and magnitudes of corrugation, as compared to the available experimental estimates \cite{Popov2012} and LMP2 data \cite{Constantinescu2013}. Therefore, such an approach is appropriate when it is needed to study the effects of structural defects or edges on in-plane relative motion or potential energy surfaces of the layers.

In the specific case of graphene and graphite, the PBE-D2 functional is found to work very well. Further improvement is achieved by using for carbon the dispersion coefficient $C_6^\mathrm{C}= 1.1725$ $\mathrm{J}\cdot \mathrm{nm}^6/\mathrm{mol}$ and van der Waals radius $R_0^\mathrm{C}=1.4868$~\AA.

As for the comparison of the calculation results for bilayer and bulk materials, it is found that for most of the considered functionals (PBE-D2, PBE-D3, PBE-D3(BJ), optPBE-vdW and vdW-DF2) the  interaction of non-adja-cent layers in bulk has a much smaller effect on characteristics of the potential energy surface than the variation of the equilibrium interlayer distance.

\section*{Acknowledgments}
Authors acknowledge the computational time on the Multipurpose Computing Complex NRC ``Kurchatov Institute". AL and AP acknowledge the Russian Foundation for Basic Research (Grant 16-52-00181). IL acknowledges the financial support from Grupos Consolidados UPV/ EHU del Gobierno Vasco (IT578-13) and EU-H2020 project ``MOSTOPHOS" (n. 646259). 

\section*{References}
\bibliography{rsc}

\end{document}